\UseRawInputEncoding
\documentclass[twocolumn]{aastex631}

\usepackage{graphicx}
\usepackage{longtable}
\usepackage{multirow}
\usepackage{rotating}
\usepackage{natbib}
\usepackage{amsmath}
\usepackage{mathtools}
\usepackage{color}
\usepackage{xspace}
\usepackage{enumitem}
\usepackage{txfonts}
\usepackage{xfrac}

\newcommand{\lsim}{\;\lower.6ex\hbox{$\sim$}\kern-7.75pt\raise.65ex\hbox{$<$}\;}
\newcommand{\gsim}{\;\lower.6ex\hbox{$\sim$}\kern-7.75pt\raise.65ex\hbox{$>$}\;}
\newcommand{\gl}{\;\lower.6ex\hbox{$<$}\kern-7.75pt\raise.65ex\hbox {$>$}\;}

\graphicspath{{./}{figures/}}

\submitjournal{ApJ}

\shorttitle{The SFR of the Milky Way as seen by \textit{Herschel}}

\shortauthors{Elia et al.}

\begin{document}
\newcommand{\sfrtot}{1.7}
\newcommand{\esfrtot}{0.6}
\newcommand{\sfrtoti}{1.5}
\newcommand{\esfrtoti}{0.5}
\newcommand{\sfrtoto}{0.3}
\newcommand{\esfrtoto}{0.1}
\newcommand{\sfrtotiperc}{84}
\newcommand{\sfrtotoperc}{16}
\newcommand{\nwithdist}{29880}
\newcommand{\nwithnodist}{5532}
\newcommand{\contrsimulmin}{0.20}
\newcommand{\contrsimulmax}{0.23}
\newcommand{\contrsimulave}{0.22}
\newcommand{\contrsimulsig}{0.01}
\newcommand{\econtrsimulave}{0.08}
\newcommand{\econtrsimulsig}{0.00}
\newcommand{\contrsimulmed}{0.13}
\newcommand{\contrsimulavg}{0.19}
\newcommand{\contrsimulcmz}{0.17}
\newcommand{\econtrsimulcmz}{0.06}
\newcommand{\sfrtottot}{2.0}
\newcommand{\esfrtottot}{0.7}
\newcommand{\contrsimulthirteenkpc}{0.57}
\newcommand{\sfrinnerlonperc}{90}
\newcommand{\sfrlongmorea}{0.11}
\newcommand{\esfrlongmorea}{0.04}
\newcommand{\sfrlongmoreb}{0.12}
\newcommand{\esfrlongmoreb}{0.05}
\newcommand{\sfryusef}{0.04}
\newcommand{\esfryusef}{0.02}
\newcommand{\sfrimmer}{0.08}
\newcommand{\esfrimmer}{0.03}
\newcommand{\sfrdns}{0.12}
\newcommand{\esfrdns}{0.05}
\newcommand{\sfrhess}{0.04}
\newcommand{\esfrhess}{0.02}
\newcommand{\sfrbarnes}{0.06}
\newcommand{\esfrbarnes}{0.02}
\newcommand{\slopessfrken}{-0.11}
\newcommand{\eslopessfrken}{0.02}
\newcommand{\slopessfrngc}{-0.20}
\newcommand{\eslopessfrngc}{0.01}
\newcommand{\slopessfrlee}{-0.25}
\newcommand{\eslopessfrlee}{0.05}
\newcommand{\slopessfrgal}{-0.28}
\newcommand{\eslopessfrgal}{0.01}
\newcommand{\offsetssfrgal}{-0.30}
\newcommand{\eoffsetssfrgal}{0.16}
\newcommand{\scalelengthsfr}{1.55}
\newcommand{\cumfifty}{5.8}
\newcommand{\cumninety}{9.2}
\newcommand{\cumninetynine}{13.4}
\newcommand{\cumrgalfour}{15}
\newcommand{\cumrgalthree}{7}
\newcommand{\cumrgalthreecomplement}{93}
\newcommand{\sfrmorethantwelve}{0.45}
\newcommand{\sfrmorethantwelveopposite}{0.60}
\newcommand{\sfrmorethantwelvediff}{0.15}
\newcommand{\sfrmorethantwelveperc}{8}
\newcommand{\sfrtottotcorrectedmorethantwelve}{2.13}
\newcommand{\plawfitcoeff}{5.6}
\newcommand{\eplawfitcoeff}{1.4}
\newcommand{\plawfitcoeffexp}{-7}
\newcommand{\plawfitcoexp}{0.74}
\newcommand{\eplawfitcoexp}{0.03}
\newcommand{\firstpeaklonmin}{336}
\newcommand{\firstpeaklonmax}{339}
\newcommand{\firstpeaklatmin}{-0.2}
\newcommand{\firstpeaklatmax}{0.0}
\newcommand{\secondpeaklonmin}{27}
\newcommand{\secondpeaklonmax}{30}
\newcommand{\secondpeaklatmin}{0.0}
\newcommand{\secondpeaklatmax}{0.2}
\newcommand{\thirdpeaklonmin}{18}
\newcommand{\thirdpeaklonmax}{21}
\newcommand{\thirdpeaklatmin}{-0.2}
\newcommand{\thirdpeaklatmax}{0.0}

\newcommand{\ksn}{1.14}
\newcommand{\eksn}{0.07}
\newcommand{\scpower}{-3}
\newcommand{\ksa}{1.39}
\newcommand{\eksa}{0.29}

\title{The Star Formation Rate of the Milky Way as seen by \textit{Herschel}}

\correspondingauthor{Davide Elia}
\email{davide.elia@inaf.it}

\author[0000-0002-9120-5890]{D. Elia}
\affil{INAF - Istituto di Astrofisica e Planetologia Spaziali, Via Fosso del Cavaliere 100, I-00133 Roma, Italy}

\author[0000-0002-9826-7525]{S. Molinari}
\affil{INAF - Istituto di Astrofisica e Planetologia Spaziali, Via Fosso del Cavaliere 100, I-00133 Roma, Italy}

\author[0000-0003-1560-3958]{E. Schisano}
\affil{INAF - Istituto di Astrofisica e Planetologia Spaziali, Via Fosso del Cavaliere 100, I-00133 Roma, Italy}

\author[0000-0002-0294-4465]{J. D. Soler}
\affil{INAF - Istituto di Astrofisica e Planetologia Spaziali, Via Fosso del Cavaliere 100, I-00133 Roma, Italy}

\author[0000-0003-0709-708X]{M. Merello}
\affil{Departamento de Astronom\'{\i}a, Universidad de Chile, Casilla 36-D, Correo Central, Santiago, Chile}

\author[0000-0001-5400-7214]{D. Russeil}
\affil{Aix Marseille Univ., CNRS, CNES, LAM, Marseille, France}

\author{M. Veneziani}
\affil{Science \& Technology Corporation, Olof Palmestraat 14, 2616 LR Delft, The Netherlands}

\author[0000-0001-9509-7316]{A. Zavagno}
\affil{Aix Marseille Univ., CNRS, CNES, LAM, Marseille, France}
\affil{Institut Universitaire de France, Paris, France}

\author[0000-0002-6296-8960]{A. Noriega-Crespo}
\affil{Space Telescope Science Institute, 3700 San Martin Dr., Baltimore, MD 21218, USA}

\author[0000-0002-1162-7947]{L. Olmi}
\affil{INAF - Osservatorio Astrofisico di Arcetri, Largo E. Fermi 5, 50125, Firenze, Italy}

\author[0000-0002-3597-7263]{M. Benedettini}
\affil{INAF - Istituto di Astrofisica e Planetologia Spaziali, Via Fosso del Cavaliere 100, I-00133 Roma, Italy}

\author[0000-0002-0472-7202]{P. Hennebelle}
\affil{Universit\'e Paris-Saclay, Universit\'e Paris Cit\'e, CEA, CNRS, AIM, 91191, Gif-sur-Yvette, France}

\author[0000-0002-0560-3172]{R. S. Klessen}
\affil{Universit\"at Heidelberg, Zentrum f\"ur Astronomie, Institut f\"ur theoretische Astrophysik,  Albert-Ueberle-Str. 2, D-69120 Heidelberg, Germany}
\affil{Universit\"at Heidelberg, Interdiszipli\"ares Zentrum f\"ur Wissenschaftliches Rechnen, Im Neuenheimer Feld 205, D-69120 Heidelberg, Germany}

\author[0000-0003-1014-3390]{S. Leurini}
\affil{INAF - Osservatorio Astronomico di Cagliari, Via della Scienza 5, I-09047 Selargius (CA), Italy} 

\author[0000-0002-5158-243X]{R. Paladini}
\affil{Infrared Processing Analysis Center, California Institute of Technology, Pasadena, CA 91125, USA}

\author[0000-0001-7852-1971]{S. Pezzuto}
\affil{INAF - Istituto di Astrofisica e Planetologia Spaziali, Via Fosso del Cavaliere 100, I-00133 Roma, Italy}

\author[0000-0003-1665-6402]{A. Traficante}
\affil{INAF - Istituto di Astrofisica e Planetologia Spaziali, Via Fosso del Cavaliere 100, I-00133 Roma, Italy}

\author[0000-0002-5881-3229]{D. J. Eden}
\affil{Astrophysics Research Institute, Liverpool John Moores University, IC2, Liverpool Science Park, 146 Brownlow Hill, Liverpool, L3 5RF, UK}
\affil{Armagh Observatory and Planetarium, College Hill, Armagh BT61 9DB, UK}

\author[0000-0002-5236-3896]{P. G. Martin}
\affil{Canadian Institute for Theoretical Astrophysics, University of Toronto, 60 St. George Street, Toronto, ON M5S 3H8, Canada}

\author[0000-0001-6113-6241]{M. Sormani}
\affil{Universit\"at Heidelberg, Zentrum f\"ur Astronomie, Institut f\"ur theoretische Astrophysik,  Albert-Ueberle-Str. 2, D-69120 Heidelberg, Germany}

\author[0000-0001-8239-8304]{A. Coletta}
\affil{INAF - Istituto di Astrofisica e Planetologia Spaziali, Via Fosso del Cavaliere 100, I-00133 Roma, Italy}
\affil{Dipartimento di Fisica, Sapienza Universit\'a di Roma, Piazzale Aldo Moro 2, I-00185, Rome, Italy}

\author[0000-0002-2636-4377]{T. Colman}
\affil{Universit\'e Paris-Saclay, Universit\'e Paris Cit\'e, CEA, CNRS, AIM, 91191, Gif-sur-Yvette, France}

\author[0000-0002-6482-8945]{R. Plume}
\affil{Department of Physics \& Astronomy, University of Calgary, 2500 University Dr. NW, Calgary, AB, T2N1N4, Canada}

\author[0000-0003-1975-6310]{Y. Maruccia}
\affil{INAF - Istituto di Astrofisica e Planetologia Spaziali, Via Fosso del Cavaliere 100, I-00133 Roma, Italy}

\author[0000-0002-2974-4703]{C. Mininni}
\affil{INAF - Istituto di Astrofisica e Planetologia Spaziali, Via Fosso del Cavaliere 100, I-00133 Roma, Italy}

\author[0000-0001-7680-2139]{S. J. Liu}
\affil{INAF - Istituto di Astrofisica e Planetologia Spaziali, Via Fosso del Cavaliere 100, I-00133 Roma, Italy}








\begin{abstract}
We present a new derivation of the Milky Way's current star formation rate (SFR) based on the data of the Hi-GAL Galactic plane survey. We estimate the distribution of the SFR across the Galactic plane from the star-forming clumps identified in the Hi-GAL survey and calculate the total SFR from the sum of their contributions. The estimate of the global SFR amounts to $\sfrtottot \pm \esfrtottot$~M$_{\odot}$~yr$^{-1}$, of which $\sfrtot \pm \esfrtot$~M$_{\odot}$~yr$^{-1}$ coming from clumps with reliable heliocentric distance assignment. This value is in general agreement with estimates found in the literature of last decades. The profile of SFR density averaged in Galactocentric rings is found to be qualitatively similar to others previously computed, with a peak corresponding to the Central Molecular Zone and another one around Galactocentric radius $R_\mathrm{gal} \sim 5$~kpc, followed by an exponential decrease as $\log(\Sigma_\mathrm{SFR}/[\mathrm{M}_\odot~\mathrm{yr}^{-1}~\mathrm{kpc}^{-2}])=a\,R_\mathrm{gal}/[\mathrm{kpc}]+b $, with $a=\slopessfrgal \pm \eslopessfrgal$. In this regard, the fraction of SFR produced within and outside the Solar circle is \sfrtotiperc\% and \sfrtotoperc\%, respectively; the fraction corresponding to the far outer Galaxy ($R_\mathrm{gal} > 13.5$~kpc) is only 1\%. We also find that, for $R_\mathrm{gal}>3$~kpc, our data follow a power law as a function of density, similarly to the Kennicutt-Schmidt relation. Finally, we compare the distribution of the SFR density across the face-on Galactic plane and those of median parameters, such as temperature, luminosity/mass ratio and bolometric temperature, describing the evolutionary stage of Hi-GAL clumps. We found no clear correlation between the SFR and the clump evolutionary stage.

\end{abstract}


\section{Introduction}\label{intro}

The star formation rate (SFR) is a widely used parameter to globally characterize galaxies \citep[see][and references therein]{ken12}. In the absence of small-scale description of the star forming activity in external galaxies, their SFR expresses the fundamental relationship between the stellar mass and the gas reservoir on which our understanding of galaxy formation and evolution is based \citep[see, e.g.,][]{mck07,kru14}.”

Several estimates of the SFR have been computed for the Milky Way as well. Considering, in principle, the total mass of available molecular gas and the free-fall time in typical conditions of clouds in the Galaxy would lead to a prediction of a global SFR of 300~M$_{\odot}$~yr$^{-1}$ \citep[reducible to 46~M$_{\odot}$~yr$^{-1}$ if only bound clouds are considered,][]{eva21}. However, the actual estimates based on direct star formation indicators converge towards a value of a few M$_{\odot}$~yr$^{-1}$ \citep[see][and references therein]{lic15}.
As shown by \citet{fra19} and \citet{boa20}, this is a typical SFR for galaxies with characteristics similar to those of the Milky Way, while in different galaxies values up to a few tens of M$_{\odot}$~yr$^{-1}$ can be reached \citep[see, e.g.,][]{cho11,mut11}.

\begin{table*}[t!]
\begin{center}
\caption{Star formation rate (SFR) estimates for the Milky Way} \label{sfrliterature}
\begin{tabular}{lcl}
\hline
\hline
Method & SFR & Reference\\

        & M$_\odot$~yr$^{-1}$     &     \\
\hline
\decimals
Ionization rate from radio free-free & 0.35\tablenotemark{a} & \citet{smi78}\\
Ionization rate from radio free-free & $2.0 \pm 0.6$\tablenotemark{a} & \citet{gue82}\\
Ionization rate from radio free-free & $1.6 \pm 0.5$\tablenotemark{a} & \citet{mez87}\\
Ionization rate from [N \textsc{ii}] 205~$\mu$m (COBE)& $2.6 \pm 1.3$\tablenotemark{a} & \citet{ben94}\\
Ionization rate from [N \textsc{ii}] 205~$\mu$m (COBE) & $2.0 \pm 1.0$\tablenotemark{a} & \citet{mck97}\\
O/B Star Counts & $1.8 \pm 0.6$\tablenotemark{a} & \citet{ree05}\\
Nucleosynthesis from $^{26}$Al (INTEGRAL) & $2.0 \pm 1.2$\tablenotemark{a} & \citet{die06}\\
Continuum emission at 100~$\mu$m (COBE) & $1.9 \pm 0.8\tablenotemark{a}$ & \citet{mis06}\\
Ionization rate from microwave free-free (WMAP) & $2.4 \pm 1.2$\tablenotemark{a} & \citet{mur10}\\
YSO counts (\textit{Spitzer}) & $1.1 \pm 0.4$\tablenotemark{a} & \citet{rob10}\\
YSO counts (MSX) & $1.8 \pm 0.3$ & \citet{dav11}\\
Combination of literature values & $1.9\pm 0.4~$ & \citet{cho11}\\
Continuum emission at 70~$\mu$m (\textit{Herschel}) & $2.1\pm 0.4~$ & \citet{nor13}\\
Combination of literature values & $1.65\pm 0.19~$ & \citet{lic15}\\
FIR clump counts (\textit{Herschel}) & $\sfrtottot\pm\esfrtottot$ & This work\\
\hline
\end{tabular}
\end{center}
\tablenotetext{a}{This value is not taken from the original article, but was re-scaled by \citet{cho11} to normalize all literature results to the same IMF (see text).}
\end{table*}

State-of-the-art estimates of the SFR, summarized in Table~\ref{sfrliterature}, provide an illustration of the variations found using different methods. The previous compilation, presented in \citet[][their Table~1]{cho11}, renormalized some of the archival values from the literature, to conform all of them to the same initial mass function \citep[IMF,][]{kro03,ken09}, as also recommended by \citet{dav11}. 

\citet{mis06}, using a direct relation between COBE/DIRBE $100~\mu$m emission and SFR already adopted for external galaxies, provided a value of 2.7~M$_\odot$~yr$^{-1}$ for the Milky Way's SFR.
\citet{rob10} estimated a SFR from 0.68 to 1.45~M$_\odot$~yr$^{-1}$, starting from young stellar objects (YSOs) revealed in the \textit{Spitzer}/IRAC GLIMPSE Galactic plane survey, and using model spectral energy distributions (SEDs) to predict the brightness and color of the synthetic YSOs at different wavelengths.

More recent results do not use new data sets, but are derived by reconsidering values existing in the literature: \citet{cho11} determine a SFR of $1.9\pm 0.4~$M$_\odot$~yr$^{-1}$ by combining and re-normalizing literature results from 1978 to 2011. \citet{lic15} derived a SFR of $1.65\pm 0.19~$M$_\odot$~yr$^{-1}$ by statistically combining the prior measurements of this quantity in the literature through a hierarchical Bayesian method. 

An independent estimate of the SFR can be given through a technique which uses far-infrared \textit{Herschel}\footnote{\textit{Herschel} is an ESA space observatory with science instruments provided by European-led Principal Investigator consortia and with important participation from NASA.} satellite \citep{pil10} data, and in particular photometric observations carried out by its PACS \citep[at 70 and 160~$\mu$m,][]{pog10} and SPIRE \citep[at 250, 350 and 500~$\mu$m,][]{gri10} cameras.
\citet{ven13} developed such method that starts from the compact sources detected in \textit{Herschel} maps. They generally correspond to unresolved clumps that already host active star formation, or, alternatively, are starless but show conditions for future activity. By considering the former category of sources, \citet{ven13} evaluated the contribution of each individual clump to the total SFR in two $2^\circ \times 2^\circ$-tiles of Hi-GAL \citep[\textit{Herschel} InfraRed Galactic Plane Survey,][]{mol10a}, namely the Open Time Key Project for unbiasedly surveying the whole Galactic plane with PACS and SPIRE. For comparison, they also derived the SFR of the same two regions through the method of \citet{li10} based on assuming a direct relation between SFR and 70~$\mu$m emission. Subsequently, by assuming that these two fields are representative of the entire Milky Way, \citet{nor13} extrapolated for the entire Galactic SFR an estimate of $2.1\pm 0.4~$M$_\odot$~yr$^{-1}$.

Other works also discussed the profile of the Milky Way SFR as a function of the Galactocentric distance $R_\mathrm{gal}$, both to quantify the ability of different zones of our Galaxy to form stars, and to make a comparison with external galaxies, for which this relation is observed. A typical approach is to study the SFR density, averaged in concentric Galactic rings, as a function of the middle radius of the ring \citep[e.g.,][]{gue82,por99,ken12,lee16}. It is generally seen that the profile of this quantity has an absolute maximum in correspondence of the Central Molecular Zone (CMZ), another local maximum around at $R_\mathrm{gal}\sim 4-5$~kpc (with a dip in the middle, centered at $R_\mathrm{gal}\sim 2$~kpc), and an exponential decrease at larger radii.

In this paper, by extending the method of \citet{ven13}, refined by \citet{ven17}, to the clump catalog of the entire Hi-GAL survey, we present a direct and self-consistent estimate of the SFR for the whole Milky Way, illustrating both the global SFR and its distribution across the Galactic plane.

In Section~\ref{sfrest} we show how Hi-GAL protostellar clump properties can be used to obtain an estimate of the current Galactic SFR. In Section~\ref{distr2d} we discuss how the SFR is distributed across the Milky Way, adopting both the $[\ell,b]$ and the pole-on perspective. We also make a test of the Kennicutt-Schmidt law with our data, and investigate possible links between local SFR and clump average evolutionary stage. Finally, in Section~\ref{summary} we draw our conclusions.

\section{The \textit{Herschel}-based Milky Way's SFR estimation}\label{sfrest}

%
\subsection{Source selection}\label{sourcesel}

We calculated the Galactic SFR using the physical properties of clumps identified in the Hi-GAL survey \citep{eli21}. These objects were identified and cataloged as follows.

The detection and 5-band photometry of Hi-GAL compact sources (i.e. unresolved or poorly resolved) were performed through the CuTEx algorithm \citep{mol11} in the five Hi-GAL bands, to obtain the single-band catalogs presented by \citet{mol16a} and Molinari et al. (in prep.). 

Starting from these lists, \citet{eli21} obtained a band-merged catalog, from which only spectral energy distributions (SEDs) with a shape eligible for a modified black body fit in the range from 160 to 500~$\mu$m were selected \citep[further details can be also found in][]{eli13,eli17}. 

Heliocentric distances were estimated by \citet{meg21} in three steps: $i$) by extracting the most reliable velocity component along the line of sight from available spectroscopic surveys of the Galactic plane; $ii$) by applying the rotation curve of \citet{rus17} and solving the near/far distance ambiguity towards the inner Galaxy by considering H\textsc{i} absorption; and $iii$) by complementing this information with independent estimates such as H\textsc{ii} region distances or maser parallaxes. The medians of the absolute and relative error on distances of \citet{meg21} are 0.54~kpc and $16\%$, respectively.

As pointed out in \citet{eli17,eli21}, the Hi-GAL compact sources are distributed along a wide range of heliocentric distances and consequently achieve a variety of physical sizes corresponding to different kinds of structures: from single cores to larger overdensities hosting a more complex but unresolved morphology.
In particular, the majority of Hi-GAL sources fulfill the definition of clump based on physical diameter \citep[$0.3 < D < 3$~pc,][]{ber07}. 

As a rough criterion to classify the clumps in star forming vs quiescent (designated for brevity as protostellar and starless, respectively), \citet{eli21} adopted the availability/lack of a detection at 70~$\mu$m, respectively. 

The modified black body fit provided temperatures for the clump SEDs and, for cases with an available distance estimate, also the mass and the bolometric luminosity. The mass was used to further classify the starless clumps as gravitationally bound (called pre-stellar) vs unbound, by using a gravitational stability criterion based on the so-called third Larson's relation \citep{lar81}. For pre-stellar sources the luminosity was estimated from the whole integral of the best-fitting modified black body. For protostellar sources, generally showing an excess of emission at $\lambda<70\,\mu$m with respect to the modified black body \citep[e.g.,][]{dun08,gia12,eli17}, the luminosity  is estimated by considering the sum of the integrals of the observed SED for $\lambda<160\,\mu$m and of the best-fitting modified black body for $\lambda \geq 160\,\mu$m, respectively. 

The same considerations made above about the physical size can be extended to the other distance-dependent observables mass and luminosity. Their ranges of variability extend over several orders of magnitude not only due to intrinsic differences among objects (that can be appreciated among sources located at the same distance), but also - and above all - due to the wide range of underlying distances, as figures~9 and~13 of \citet{eli21} clearly illustrate. More quantitatively, 99\% of Hi-GAL clumps of \citet{eli21} provided with a distance estimate have masses ranging from 0.1 to $10^4$~M$_\odot$, and luminosities ranging from 0.1 to $10^5$~L$_\odot$.

To compute the SFR through the method described in Section~\ref{sfrcalc}, first we consider all \nwithdist~protostellar clumps provided with a heliocentric distance. Subsequently, in Section~\ref{nodist} we evaluate and suggest a reasonable additional term which accounts for the contribution by further \nwithnodist~protostellar clumps lacking a distance determination.


\subsection{Computation of the SFR}\label{sfrcalc}
\label{method}
To determine the SFR we followed the method described by \citet{ven13} and refined by \citet{ven17}. 

In short, the contribution of each protostellar source to the total SFR is computed through theoretical evolutionary tracks reported in the bolometric luminosity vs mass diagram \citep{mol08,bal17b,eli21}. At each time step, together with the clump mass-luminosity pair, the mass of the internal protostar being formed is also computed. At the end point of the theoretical track both the final mass of the star and the corresponding total elapsed time are known. The ratio between these two quantities constitutes the SFR contribution from the given clump.

The evolutionary tracks were initially obtained by \citet{mol08} for the case of an object forming a single central star but, especially at higher masses, we actually deal with clumps possibly hosting the formation of entire protoclusters \citep[e.g.][]{bal17a}. \citet{ven17} took into consideration this aspect by applying a Monte Carlo procedure to an evolutionary model of turbulent cores to account for the wide multiplicity of sources produced during the clump collapse. We implemented this refinement in this work as well \citep[for details, see][]{ven17}.

Furthermore, given the wide range of clump heliocentric distances quoted in the Hi-GAL catalog \citep{eli17,eli21}, a distance bias might be expected to affect the estimate of physical parameters for these objects. \citet{bal17a} proposed a method to evaluate this possible bias by simulating the appearance of  nearby star forming regions observed by the \textit{Herschel} Gould Belt Survey \citep{and10} if they were moved to larger distances (from 0.75 to 7~kpc), comparable to those typically found in the Hi-GAL survey. Subsequently, in \citet{bal17b} this approach was applied, in particular, to the estimation of the SFR. For three nearby regions they compared the SFR obtained from YSO counts \citep[based on an updated version of Equation~1 of][]{lad10} to that estimated through the method of \citet{ven17} from the clumps detected in the \textit{Herschel} maps ``displaced'' at different virtual distances. They found that the two estimates remain consistent with each other within a factor 2 at various probed distances.


Based on these premises we feel confident to adopt the method of \citet{ven17}. As mentioned above, the output of the algorithm we used is the contribution of each clump $\Sigma_\textrm{cl}$ to the global SFR. This quantity depends on the evolutionary track in the $L/M$ plot, which, in turn, is defined by the initial clump mass $M_\textrm{cl}$. Although not explicitly stated in the original articles, this algorithm associates the final star mass to the input clump mass by interpolating the locus of final masses of known evolutionary tracks with a power law. Here we explicitly report the corresponding formula we used:
\begin{equation}
   \textrm{SFR}_\textrm{cl}=(\plawfitcoeff \pm \eplawfitcoeff) \times 10^{\plawfitcoeffexp}~ (M_\textrm{cl}/\textrm{M}_{\odot})^{\plawfitcoexp \pm \eplawfitcoexp}~\textrm{M}_{\odot}~\textrm{yr}^{-1}. 
   \label{eqsfr}
\end{equation}

We propose this relation hereafter as an operational prescription for an immediate application of the method of \citet{ven17}.

The global SFR, estimated by adding up all these contributions, amounts to $\sfrtot \pm \esfrtot$~M$_{\odot}$~yr$^{-1}$. The uncertainty is given by the error propagation, combining in quadrature the uncertainties affecting the clump masses and the parameters appearing in Equation~\ref{eqsfr}.

Adopting a classification of inner vs outer Galaxy delimited by the Solar circle with a radius of~8.34~kpc \citep[as in][]{meg21,eli21}, it is possible to give separately the contributions to the global SFR from these two zones: \sfrtoti~(i.e. the \sfrtotiperc\% of the total) and \sfrtoto~M$_{\odot}$~yr$^{-1}$ (\sfrtotoperc\%), respectively.

\subsubsection{Contribution from sources devoid of a distance estimate}\label{nodist}

\begin{figure}[!t]
\includegraphics[width=0.47\textwidth]{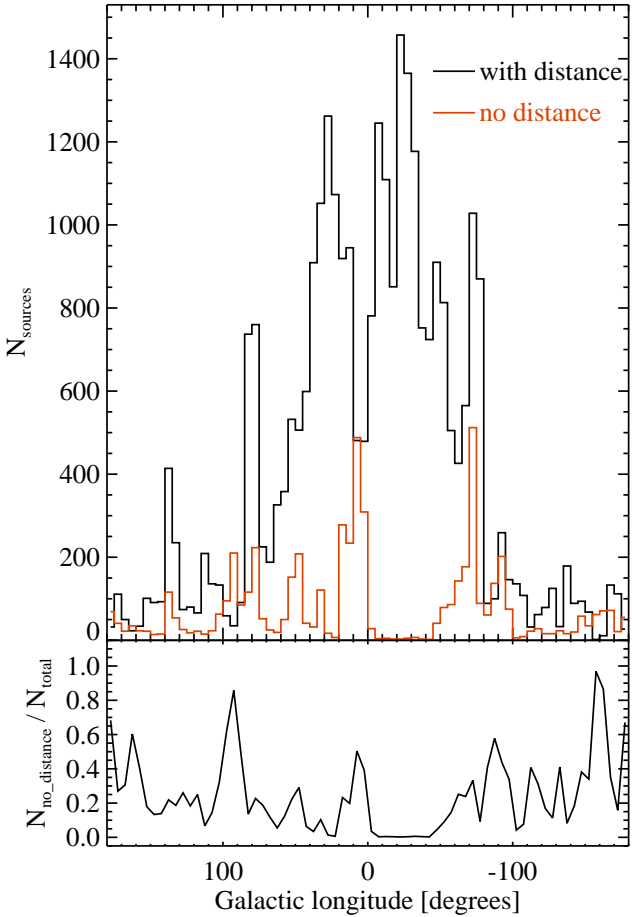}
\caption{\textit{Top}: distributions of Hi-GAL protostellar clumps with and without a heliocentric distance estimate (black and red histograms, respectively) in $5^\circ$-wide bins of Galactitc longitude. \textit{Bottom}: number ratio between protostellar clumps without a distance estimate and the total number. in the same longitude bins as in the top panel.}
\label{glondistanceless}
\end{figure}

It is to notice that
the SFR estimate reported above should be intended as a lower limit, since the input data, namely the protostellar clump list, is incomplete for the reasons described in the following. First, since the mass completeness in the Hi-GAL catalog is a function of the heliocentric distance \citep{eli17}, faint and/or far and/or cold protostellar clumps might remain undetected  at all. We address this issue in Appendix~\ref{complappendix}. Second, protostellar clumps  undetected at 70~$\mu$m (for the same reasons written above), but detected at the other ones, result misclassified as starless and are not involved in the SFR calculation. In fact, in the literature examples can be found of spectroscopic signatures of active star formation detected in 70~$\mu$m-dark clumps \citep[e.g.,][]{tra17}. Third, \nwithnodist~sources classified as protostellar in the Hi-GAL catalog are not provided with a distance, and consequently not involved in our previous calculation (see Figure~\ref{glondistanceless}). 

\begin{figure}[t]
\includegraphics[width=0.47\textwidth]{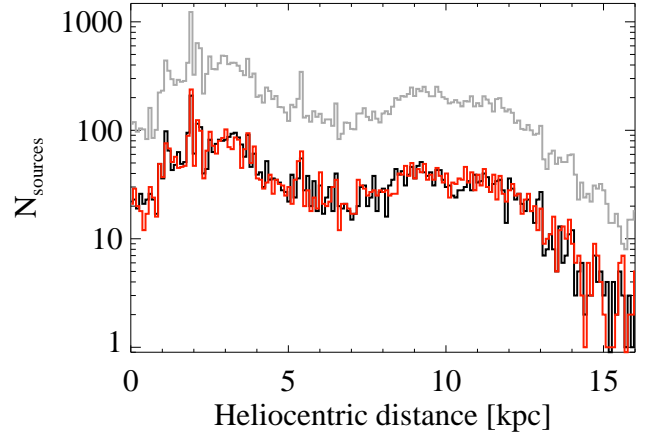}
\caption{Two examples (black and red histograms, respectively, in bins of 100~pc) of realizations of heliocentric distance distributions to be assigned, as a test, to Hi-GAL protostellar clumps devoid of a distance estimate, compared with the distance distribution for sources provided with a distance estimate \citep[gray histogram,][]{eli21}.}
\label{distdistr}
\end{figure}

Regarding the last point, it is possible to give an estimate of the missing contribution from this subsample by simulating a plausible realization of distances for sources whose distance is unknown. In this way, masses can be obtained also for them, to be inputted in the SFR calculation. 

We simulated a set of \nwithnodist~distances whose distribution follows the one of the protostellar sources provided with a distance estimate. Then we randomly assigned each simulated distance to a clump devoid of distance, computed mass and luminosity accordingly, and estimated the contribution to the SFR with the usual procedure. This random assignation of distances to clumps has been repeated 100 times (two of which are shown in Figure~\ref{distdistr}), to keep safe from specific effects possibly induced by a single realization. The 100 corresponding estimates of additional SFR contribution are found to be confined in the relatively narrow range \contrsimulmin~to \contrsimulmax~M$_{\odot}$~yr$^{-1}$, with an average of \contrsimulave~M$_{\odot}$~yr$^{-1}$ and a standard deviation of \contrsimulsig~M$_{\odot}$~yr$^{-1}$.
Notice that this standard deviation is expected to decrease by increasing the number of simulation as desired. The real uncertainty, instead, is dominated by the typical error bar associated to the SFR contribution evaluated for each simulation, which is $\econtrsimulave$~M$_{\odot}$~yr$^{-1}$. 

In conclusion, a reasonable estimate of the total Galactic SFR taking into account the correction calculated above is $\sfrtottot \pm \esfrtottot$~M$_{\odot}$~yr$^{-1}$.  

\begin{figure}[t]
\includegraphics[width=0.47\textwidth]{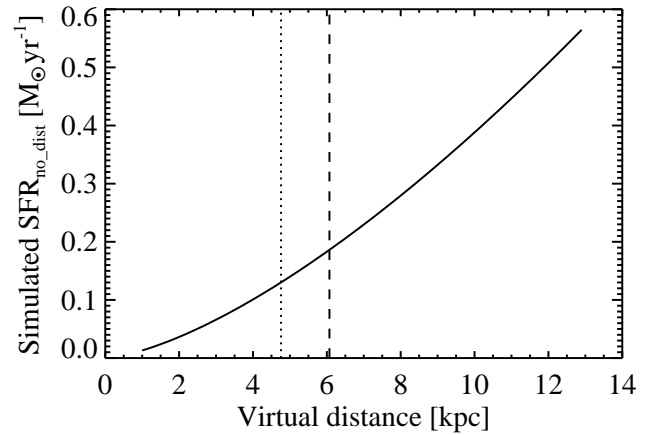}
\caption{Probable contribution to the overall SFR from sources lacking a distance estimate if they were all located at the same virtual distance. The dotted and dashed lines represent the median and the average of the real distance distribution, respectively.
}
\label{distsimul}
\end{figure}

The above test is based on the assumption that the real (unknown) distances of sources lacking this information follow the same distribution of the known ones. However, the lack of a distance estimate, i.e. the impossibility of individuating a clear and/or reliable spectral component along the line of sight of the Hi-GAL source \citep{meg21}, could be, in principle, due to specific local conditions. This could then produce a departure from the assumed behavior of distance distributions. In this respect, the most unfavorable hypothetical case would be to have all sources located at the same distance. Therefore, supposing that all sources without distance were located at a minimum and at a maximum distance, respectively, helps us to identify a range of maximum variability of their contribution to the total SFR. We therefore analyzed the behavior of such contribution for a common distance varying between 1 and 13~kpc (these limits are suggested by the distributions shown in Figure~\ref{distdistr}). From Figure~\ref{distsimul} we see that for a virtual common distance of 1~kpc the contribution to the SFR would be negligible, while for 13~kpc it would amount to \contrsimulthirteenkpc~M$_{\odot}$~yr$^{-1}$. Clearly, both scenarios (and, consequently, the corresponding contributions to SFRs) are unrealistic, but this test proves that, even in case of an extremely unfavorable distance distribution of clumps without distance assignment, the order of magnitude of the SFR increment estimated through our test would be confirmed. Finally, we notice that, assuming that all sources without known distance were placed at the median and the average of the known distance distribution, their contribution would amount to \contrsimulmed~and \contrsimulavg~M$_{\odot}$~yr$^{-1}$, respectively (see Figure~\ref{distsimul}).

A final test we propose consists in a hybrid approach between the two followed above. Indeed, the top panel of Figure~\ref{glondistanceless} suggests that the distribution in longitude of clumps lacking a distance assignment is far from being uniform and, moreover, it does not strictly follow the behavior of sources with a distance (as testified by the high degree of scatter of the ratio between the former distribution and the overall one, shown in the bottom panel). Statistically significant amounts of sources without a distance are found in correspondence of the innermost Galactic longitudes ($0^\circ \lesssim \ell \lesssim 20^\circ$), and in the direction of the Carina arm ($270^\circ \lesssim \ell \lesssim 310^\circ$). The first case can be treated with relative ease, by arbitrarily placing all the sources without distance and with $|\ell| < 20^\circ$ at the Galactic center distance of 8.34~kpc, and proceeding for the remaining sources by simulating distances distributed as the known ones. In this case, the additional term to the global SFR would amount to $\contrsimulcmz \pm \econtrsimulcmz$~M$_{\odot}$~yr$^{-1}$. 

\subsection{Comparison with the previous literature}
Adopting the value of $\sfrtottot \pm \esfrtottot$~M$_{\odot}$~yr$^{-1}$ as our final estimate for the Milky Way SFR, and comparing with values in Table~\ref{sfrliterature}, it can be seen that it is consistent, within the error bars, with all but one \citep[namely][]{smi78} estimates collected and normalized by \citet{cho11}, and with the estimates of \citet{cho11} themselves and of \citet{lic15}. Remarkably, our estimate is also perfectly consistent, within the errors, with that of \citet{nor13}, confirming that his choice to extrapolate the SFR estimate from an area of 8~sq.~deg. to the entire 720~sq.~deg was quite reasonable.

Our result, in essence, corroborates the common idea - supported by the fact that a variety of approaches leads to similar values - that the current SFR of the Milky Way lies in the range 1.5-2.5~M$_{\odot}$~yr$^{-1}$, which is consistent with the values found for spiral galaxies \citep[e.g.,][]{ken12}.


\section{The SFR across the Galactic plane}\label{distr2d}

We studied the SFR distribution across the Galactic Plane. From here on, we will only consider the contribution to SFR given by clumps whose heliocentric distance is known, for which the position in the Galaxy can be determined. 

\subsection{Mapping SFR in Galactic coordinates: the Central Molecular Zone}
\begin{figure*}[!t]
\includegraphics[width=0.97\textwidth]{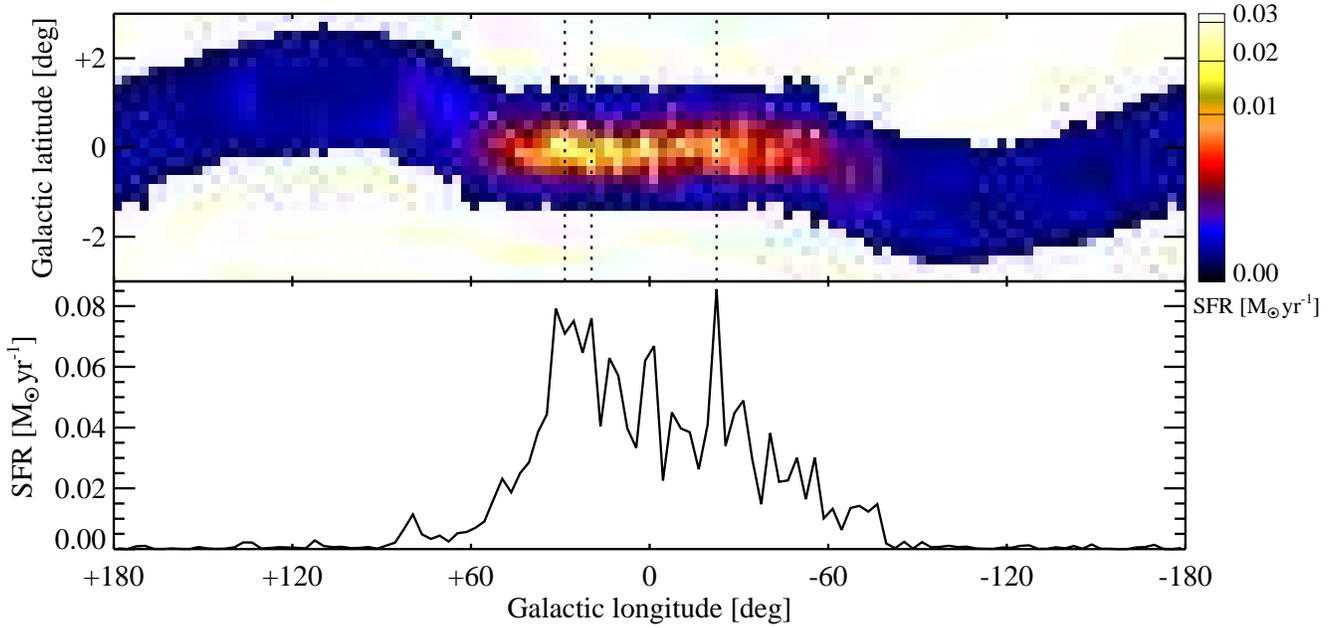}
\caption{\textit{Top}: Star formation rate (SFR) toward the Galactic plane in  $3^\circ \times 0.2^\circ$ bins in Galactic longitude and latitude, respectively. It is to notice that in the outer Galaxy the Hi-GAL latitude coverage follows the variations in the midplane position, or Galactic warp \citep{wes57}. For a better rendering of the image dynamic range, the color scale follows a power law with exponent 0.4. The dotted lines indicate the location of the three most prominent SFR peaks. \textit{Bottom}: SFR as a function of longitude, estimated from the sum of the results reported in the top panel across Galactic latitude. 
}
\label{maplbsfr}
\end{figure*}

A first view of SFR distribution in the Galactic plane can be given in the $[\ell,b]$ plane, as shown in Figure~\ref{maplbsfr}, top. It can be seen that, approximately, most of the total SFR (\sfrinnerlonperc\%) comes from the inner 120$^\circ$ (Figure~\ref{maplbsfr}, bottom). 

For the used binning of 3$^\circ$ in longitude and 0.2$^\circ$ in latitude, the three most prominent local maxima are found in pixels corresponding to the ranges $\firstpeaklonmin^\circ <\ell < \firstpeaklonmax^\circ$ and $\firstpeaklatmin^\circ < b < \firstpeaklatmax^\circ$ \citep[containing the G337.342−0.119 region,][]{jac18}, $\secondpeaklonmin^\circ <\ell < \secondpeaklonmax^\circ$ and $\secondpeaklatmin^\circ < b < \secondpeaklatmax^\circ$, and $\thirdpeaklonmin^\circ <\ell < \thirdpeaklonmax^\circ$ and $\thirdpeaklatmin^\circ < b < \thirdpeaklatmax^\circ$, respectively.

\begin{table*}[t!]
\begin{center}
\caption{Star formation rate estimates for the Central Molecular Zone } \label{sfrcmz}
\begin{tabular}{llclc}
\hline
\hline
Method  & Area boundaries & SFR$_\textrm{CMZ}$ (lit.) & Reference & SFR$_\textrm{CMZ}$ (this work)\\
    &   & M$_\odot$~yr$^{-1}$     & &M$_\odot$~yr$^{-1}$      \\
\hline
\decimals
YSO counts (\textit{Spitzer}) &  $|\ell| < 1^\circ, |b| < 10\arcmin$ & 0.14 & \citet{yus09}& $\sfryusef \pm \esfryusef$\\
Continuum emission at 60, 100~$\mu$m (IRAS) &  $|\ell| < 3^\circ, |b| < 1^\circ$ \tablenotemark{a} & 0.12 & \citet{cro11}& $\sfrdns \pm \esfrdns$\\
Continuum emission at 60, 100~$\mu$m (IRAS) &  $|\ell| < 0.8^\circ, |b| < 0.3^\circ$ & 0.08 & \citet{cro11}& $\sfrhess \pm \esfrhess$\\
YSO counts (\textit{Spitzer}) &  $|\ell| < 1.5^\circ, |b| < 0.5^\circ$ & 0.08 & \citet{imm12}& $\sfrimmer \pm \esfrimmer$\\
 Ionization rate from radio free-free &  $2.5^\circ < \ell < 3.5^\circ, |b| < 0.5^\circ$ & 0.035 & \citet{lon13}& $\sfrlongmorea \pm \esfrlongmorea$\\
  Ionization rate from radio free-free &  as above, but $|b| < 1^\circ$ for $|\ell| < 1^\circ$ & 0.06 & \citet{lon13}& $\sfrlongmoreb \pm \esfrlongmoreb$\\
Continuum emission at 24~$\mu$m (\textit{Spitzer})  &  $|\ell| < 1^\circ, |b| < 0.5^\circ$ & $0.09\pm 0.02$ & \multirow{3}{*}{\hspace{-1em}$\left.\begin{array}{l}        \\            \\              \\ \end{array}\right\rbrace$}\\
Continuum emission at 70~$\mu$m (\textit{Spitzer})  &  $|\ell| < 1^\circ, |b| < 0.5^\circ$ & $0.10\pm 0.02$ & ~~\citet{bar17}\tablenotemark{b}& $\sfrbarnes \pm \esfrbarnes$\tablenotemark{c}\\
Cont. emission at 5.8-500~$\mu$m (\textit{Spitzer}, \textit{Herschel})  &  $|\ell| < 1^\circ, |b| < 0.5^\circ$ & $0.09\pm 0.03$ & & \\
\hline
\end{tabular}
\end{center}
\tablenotetext{a}{Differently from other rectangular areas quoted in the table, this one has an elliptical shape with axes aligned along $\ell$ and $b$, and semi-axes of $3^\circ$ and $1^\circ$, respectively.}
\tablenotetext{b}{In \citet{bar11}, their Table~2, a further list of SFR$_\textrm{CMZ}$ estimates derived within the same area from monochromatic fluxes or bolometric luminosities available in the literature from 1998 to 2011 are quoted (see text). We omit them in this table for the sake of brevity.}
\tablenotetext{c}{The three SFR$_\textrm{CMZ}$ estimates by \citet{bar17} are derived within the same area in the sky through different methods; obviously, for the same area we can provide only one estimate.}
\end{table*}

Through the $[\ell,b]$ view of the Galactic SFR it is also possible to make a direct comparison with SFRs estimated in the literature for the CMZ, generally taken as a rectangular coordinate box in such plane. In Table~\ref{sfrcmz} we report a list of recent SFR$_\textrm{CMZ}$ estimates, each of them derived in a different area around the Galactic center, and the output of our method for the same area. 

We found a good agreement with the SFR$_\textrm{CMZ}$ estimates reported in \citet{cro11} and \citet{imm12}. We also found a fairly good agreement with \citet{bar17} on SFR$_\textrm{CMZ}$, which those authors estimated from monochromatic fluxes or bolometric luminosities available in the literature from 1998 to 2011, and ranging from 0.07 to 0.12~M$_{\odot}$~yr$^{-1}$, to be compared with our estimate of $\sfrbarnes\pm\esfrbarnes$~M$_{\odot}$~yr$^{-1}$.

Our results, however, appear to disagree with the SFR estimated for the CMZ in \citet{yus09} and \citet{lon13}. The SFR$_\textrm{CMZ}$ in \citet{yus09} is above our result, which is most likely due to an excessive number of early YSOs involved in the calculation, as discussed in \citet{koe15}, who quantify this overestimate in a factor of~3. Correcting by this factor, the estimate of \citet{yus09} would turn out to be consistent with our value ($\sfryusef \pm \esfryusef$~M$_{\odot}$~yr$^{-1}$) within the uncertainties.

\citet{lon13} estimated SFR$_\textrm{CMZ} = 0.035$~M$_{\odot}$~yr$^{-1}$ in the range $2.5^\circ<\ell<3.5^\circ$ and $|b|<0.5^\circ$, and 0.06~M$_{\odot}$~yr$^{-1}$ if the wider latitude range $|b|<1^\circ$ is considered in the central $|\ell|<1^\circ$. Correspondingly, we calculate SFR$_\textrm{CMZ}=0.11$ and 0.12~M$_{\odot}$~yr$^{-1}$ for these two zones, respectively. The former one is about 3~times larger than that of \citet{lon13}, and not compatible with it within the associated error bar. Of course, the estimate of \citet{lon13} is also the lowest among the estimates from the literature collected in Table~\ref{sfrcmz}. The authors indeed suggest that the choice of the assumed IMF, which did not account for possible peculiar conditions of the CMZ, could affect this estimate down to a factor~1/3.

\begin{figure*}[!t]
\plotone{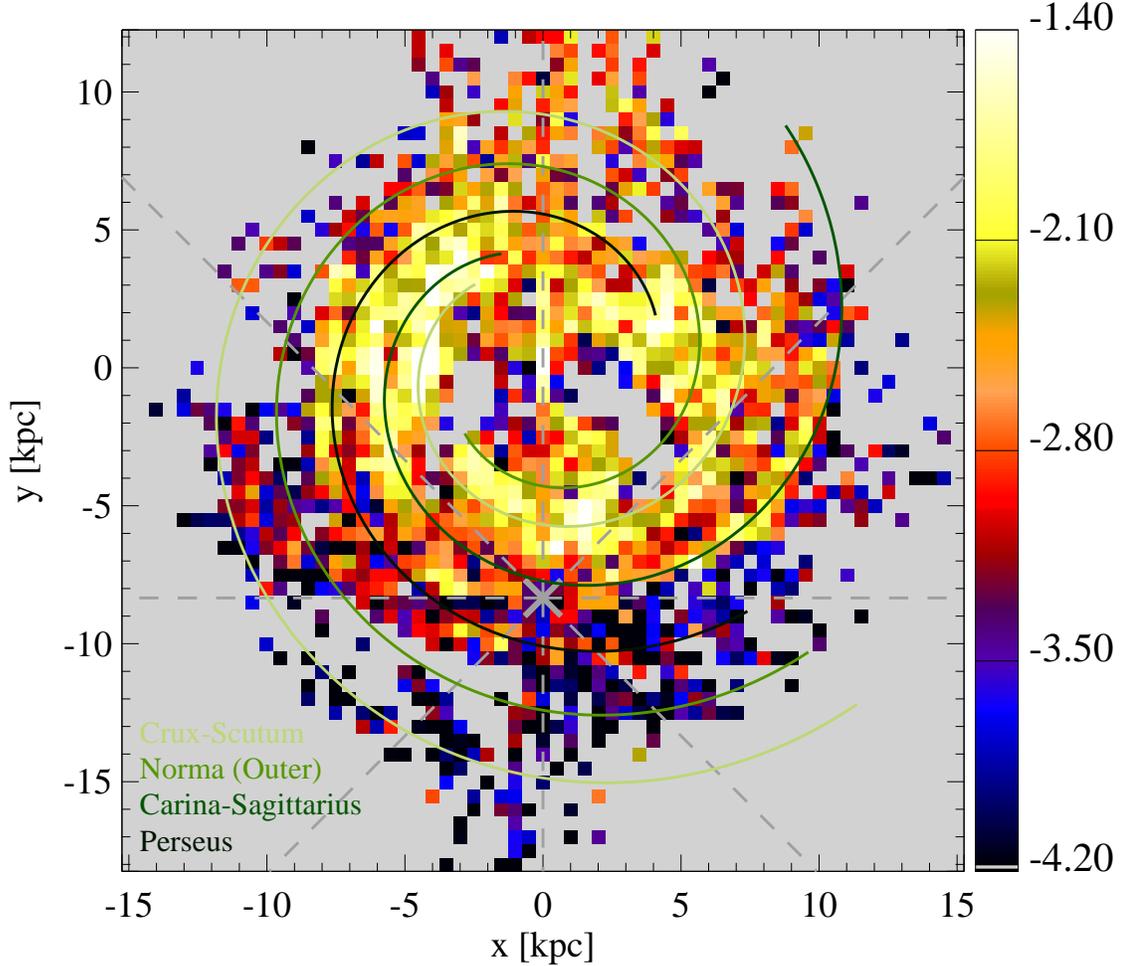}
\caption{Map of the decimal logarithm of SFR density (divided by M$_{\odot}$~yr$^{-1}$~kpc$^{-2}$), across the Galactic plane. The spatial bins are of $0.5\times 0.5$ kpc$^2$, so that the absolute SFR inside each pixel can be obtained by exponentiating the displayed logarithm, and then multiplying by a factor 4~kpc$^2$. The Galactic center is at the [0,0] position, while the gray X symbol indicates the position of the Sun. Gray dashed lines mark Galactic octants. Four spiral arms according to the prescription of \citet{hou09} are overplotted with four different tones of green; the correspondence between arms and colors is explained in the legend at the bottom left corner.}
\label{mapsfr}
\end{figure*}

\subsection{SFR Galactocentric profile}

Using the information about the heliocentric distance, a pole-on view of SFR density in the Milky Way, mapped on a grid of $0.5\times 0.5$-kpc$^2$ pixels, can be built.
This is a simplified 2-D representation of the Galactic disk which has a structure along the third dimension, an idea of which can be taken from Figure~\ref{maplbsfr}. The design of the Hi-GAL survey took into account the warp of far-infared emission already observed by IRAS towards the outer Galaxy, which however is less prominent than the Galactic warp observed in atomic gas \citep{wes57,sol22}, corresponding to Galactocentric distances not achieved by objects in our sample \citep[$R_\mathrm{gal} \gtrsim 16$~kpc,][]{vos06}.

In Figure~\ref{mapsfr} one can appreciate that, with respect to the Galactic center (to which we assign the [0,0] position), the SFR shows a certain degree of circular symmetry.

\begin{figure}[!ht]
\includegraphics[width=0.47\textwidth]{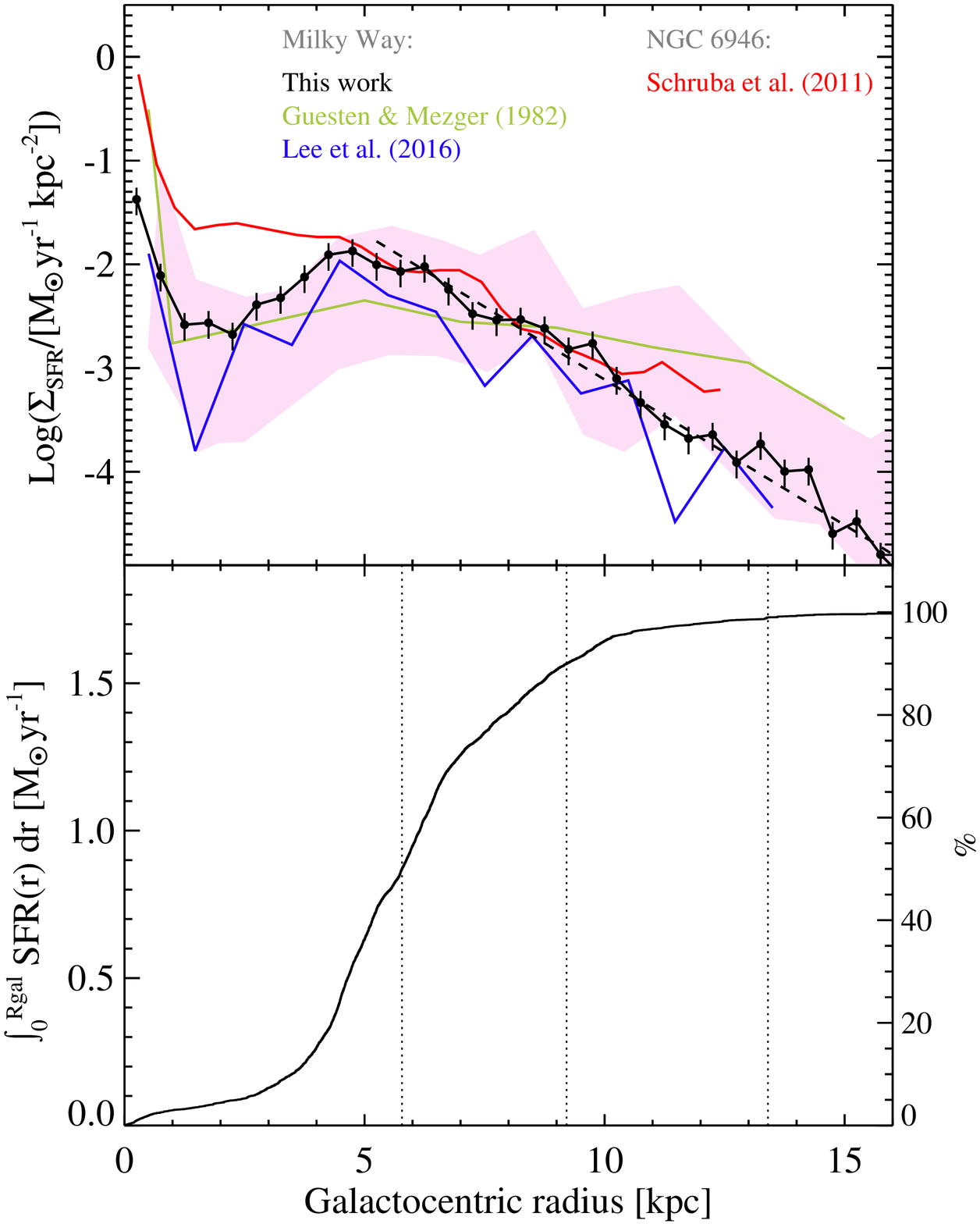}
\caption{\textit{Top}: Star formation rate (SFR) density profile estimated from the Hi-GAL observations in 0.5-kpc-wide concentric rings around the Galactic center, shown in black filled circles. Fitting the relation 
in Eq.~\ref{logsfr} to the portion at $R_\mathrm{gal}\geq 5$~kpc gives $a=\slopessfrgal \pm \eslopessfrgal$ and $b=\offsetssfrgal \pm \eoffsetssfrgal$. The fit is plotted as a black dashed line. The green line shows the results of the visible-band observations considered in \citet{gue82}, re-scaled by \citet{ken12} to the total SFR of 1.9~M$_\odot$~yr$^{-1}$ reported in \citet{cho11}.
The blue line corresponds to the estimates obtained in \citet{lee16} by combining YSO NIR photometry and cloud properties from millimeter line observations. The red line shows the SFR profile of the galaxy NGC~6946, presented in \citet{sch11} and adapted by \citet{ken12}.
The pink-shaded area is the region occupied by the models of \citet{eva22}.
\textit{Bottom}: Cumulative SFR profile as a function of the Galactocentric radius, also expressed as a percentage with respect to the total SFR ($y$-axis on the right-hand side). The vertical dotted lines indicate, from left to right, the Galactocentric radii corresponding to 50\%, 90\%, and 99\% of the total SFR, respectively.}
\label{radialsfr}
\end{figure}

It makes sense, therefore, to analyze the behavior of the SFR density $\Sigma_\mathrm{SFR}$, averaged in concentric 500-pc wide rings, vs the Galactocentric radius $R_\mathrm{gal}$.
In Figure~\ref{radialsfr}, top panel, a peak corresponding to the inner circle\footnote{To correctly compare this plot with Figure~\ref{mapsfr}, it must be considered that in this case the SFR density is obtained by dividing by the ring area, proportional to $R_\mathrm{gal}^2$ , while in Figure~\ref{mapsfr} the denominator consists of a local area extending over 0.25~kpc$^2$. Also notice that, as discussed in Section~\ref{nodist}, the distribution in longitude of sources without a distance assignment shows a remarkable concentration at $0^\circ \lesssim \ell \lesssim 20^\circ$. Assigning the distance of the Galactic center to all of them would turn out in an even higher value for this peak.} is followed by a wide dip between 1 and 4~kpc, mostly due to shortage of sources at these Galactocentric distances \citep{eli21}. This is seen also in other works (see below), and is due in part to the lack of SFR data in two lobes extending south-west and south-east of the Galactic center position over an extent of about 4~kpc (recognizable as inner void areas in Figure~\ref{mapsfr}), which can not be populated with heliocentric distances since the observed clump velocities are forbidden by the adopted rotation curve \citep{rus17}. In particular, in these areas absolute values of radial velocity are expected to be larger than 100~km~s$^{-1}$ \citep[][their Fig.~1]{ell15}, while most of sources observed in the corresponding longitude ranges show lower values. The kinematics of the Milky Way along these lines of sight is indeed generally dominated by the Galactic bar \citep{bin91,ell15, li16}. As explained by \citet{meg21} and \citet{eli21}, only in cases of small deviations, the distance of the corresponding tangent point is assigned to sources, which can therefore be involved in the calculation of the SFR. Additionally, in the longitude range corresponding to the void lobe in the fourth quadrant, \citet{meg21} found a relatively high occurrence of noise in the original spectral survey data which in many cases prevented an estimation of the radial velocity and hence of the kinematic distance. In any case, since the two void areas are oriented approximately along Galactocentric radial directions, their influence is distributed over a certain number of rings (and therefore attenuated). 

The dip in the SFR density profile is followed by another local peak at around 5~kpc. A peak in this position was predicted by \citet{kru05} as a consequence of the distribution of the molecular gas surface density, and was found also by \citet{por99} and \citet{chi01}, but at $\sim 4$~kpc. It should correspond to the so-called Molecular Ring \citep[which is, more realistically, an effect of spiral arm arrangement, see][]{dob12,rom16,miv17}, but it does not mirror exactly the peak of source number distribution found by \citet{eli17} at $R_\mathrm{gal}\simeq 6$~kpc. Finally, at larger distances, $\Sigma_\mathrm{SFR}$ definitely shows a systematic decrease. 

This global behavior is qualitatively similar to those shown by \citet{ken12} \citep[starting from the data of][]{gue82} and \citet{lee16}: an absolute maximum corresponding to the CMZ, followed by a local minimum and by another local peak at $R_\mathrm{gal}=$1 and~5~kpc, respectively, with a decreasing behavior at larger Galactocentric distances. This general behavior was successfully modeled by \citet{eva22} by taking into account $i$) a dependence of the conversion factor from CO luminosity to cloud mass on metallicity (which in turn depends on $R_\mathrm{gal}$), and $ii$) a dependence of the star formation efficiency on the virial parameter. The envelope of the six models produced by \citet[][their Fig.~3]{eva22} in the $\Sigma_\mathrm{SFR}$ vs $R_\mathrm{gal}$ diagram delimits a belt (with a vertical width of 1-2 dex) in which the curve observed by \citet{lee16} is contained. Being the latter quite similar to ours in many points (Figure~\ref{radialsfr}, top), a similar consistency is expected also for our data. In fact, our $\Sigma_\mathrm{SFR}$ curve is found, in turn, to be enclosed within the region occupied by the models of \citet{eva22}, running closer to the top of it around the 
$R_\mathrm{gal}\simeq 5$~kpc peak, and to the bottom of it at larger Galactocentric distances, in the range corresponding to the final decrease.

It is also important to compare the SFR profile found for Milky Way with that of external galaxies, to examine if some macroscopic bias is introduced when observing the Galaxy from the inside, rather than as a whole from the outside. The SFR Galactocentric profile of NGC~6946 galaxy obtained by \citet{sch11} and adapted by \citet{ken12} shows, in turn, a qualitatively similar, but flatter behavior. More recently, \citet{log19}, using H$\alpha$ emission, estimated for NGC~3994 and NGC~3995 SFR profiles which, although not quantitatively similar to that of the Milky Way, again broadly follow the same sequence of peaks and dips. Interestingly, an similar inner dip was also found by \citet{lia18} to be typical of local massive star-forming galaxies ($10.5 <\log(M/M_\odot) < 11$).

Returning to SFR density profile of the Milky Way and to its decrease for $R_\mathrm{gal}\gtrsim 5$~kpc, \citet{mis06} found a good agreement between the radial profile of SFR density collected from the literature and their model of spatial distribution of dust, stars, and gas in the Milky Way, constrained through COBE/DIRBE data. Such model has a functional form
\begin{equation}
\log(\Sigma_\mathrm{SFR}/[\mathrm{M}_\odot~\mathrm{yr}^{-1}~\mathrm{kpc}^{-2}])=a\,R_\mathrm{gal}/[\mathrm{kpc}]+b\:,
\label{logsfr}
\end{equation} 
with $a \sim -0.14$. Fitting the same relation to our data for $R_\mathrm{gal}\geq 5$~kpc, we obtain $a=\slopessfrgal \pm \eslopessfrgal$ (Figure~\ref{radialsfr}, top). 

The linear fit in the same range of Galactocentric distance for the \citet{gue82} has a much shallower slope ($a=\slopessfrken \pm \eslopessfrken$), while the more recent curve of \citet{lee16}, which shows more scattering, has a slope consistent with ours: $a=\slopessfrlee \pm \eslopessfrlee$. Finally, in the NGC~6946 case the slope shows an intermediate behavior ($a=\slopessfrngc \pm \eslopessfrngc$).

The slope of \slopessfrgal\, corresponds to an exponential scale length of \scalelengthsfr~kpc, which is considerably shorter than those typically found for the stellar count profile in the Galactic thin disk, e.g., 2.15~kpc by \citet{lic15}, 2.6~kpc by \citet[][see also further references therein]{bla16}, and 2.2~kpc by \citet{xia18}. However more recently, by using LAMOST \citep{cui12,den12} Data Release~4, and Gaia Data Release~2 \citep{gai18, lin18}, \citet{yu21} derived shorter scale lengths. In particular, they considered separately stellar populations characterized by different chemical abundances; the scale length that most closely approaches ours is that found for stars with Solar-type abundances, 1.28~kpc (but for $R_\mathrm{gal} > 8$~kpc).

In Figure~\ref{radialsfr}, bottom, the behavior of the SFR profile as a function of $R_\mathrm{gal}$ is further analyzed by means of its cumulative. Although this curve is reminiscent of the SFR density profile shown in the top panel, it additionally accounts for the increasing area of the ring at increasing $R_\mathrm{gal}$. For our data we notice three most evident changes of slope in the cumulative at around 4 (marked steepening), 7 (getting shallower), and 10~kpc (flattening), respectively. We find that the 50\%, 90\%, and 99\% of total SFR are achieved at $R_\mathrm{gal}=$~\cumfifty, \cumninety, and \cumninetynine~kpc, respectively. This means that, according to our results, half of the Galactic SFR comes from within the Molecular Ring, and just 1\% from the far outer Galaxy \citep[$R_\mathrm{gal}>13.5$~kpc, according to the definition by][]{hey98}.

\subsection{Testing the Kennicutt-Schmidt relation}\label{kssection}
The SFR density profile computed in rings of $R_\mathrm{gal}$ and plotted in the top panel of Figure~\ref{radialsfr} can be also used for checking if the SFR in the Milky Way follows the Kennicutt-Schmidt (KS) law. 

This is an empirical scaling relation \citep{sch59,ken98} generally found between star formation rate density $\Sigma_\mathrm{SFR}$ and mean gas (atomic + molecular) surface density $\Sigma_\mathrm{gas}$ in disk galaxies \citep[the atomic and molecular gas density, $\Sigma_\mathrm{H\textsc{i}}$ and $\Sigma_{\mathrm{H}_2}$, respectively, can be also considered separately, see][]{big08,ken12}. It is expressed by a power law such as 
\begin{equation}
    \Sigma_\mathrm{SFR} \propto \Sigma_\mathrm{gas}^n\,,
\end{equation}
with $n$ generally falling in the range 1~to~2, depending on the tracer used and the linear scales considered. In particular, the ``classic'' slope of \citet{ken98} is $n=1.40 \pm 0.15$, with $\Sigma_\mathrm{gas}$ expressed in units of M$_{\odot}$~pc$^{-2}$. Finally, in the literature descriptions of $\Sigma_\mathrm{SFR}$ as a power law of the single atomic ($\Sigma_\mathrm{H\textsc{i}}$), or molecular component ($\Sigma_\mathrm{H_2}$), respectively, can be also commonly found.


Whereas a vast literature is available about external galaxies following the KS relation \citep[see, e.g.,][and references therein]{ken98,mie17,orr18}, this has been poorly tested for the Milky Way, also due to well understood intrinsic difficulties related to observing the Galaxy ``from within''. 


The KS relation for a set of regions in the Southern Milky Way was explored by \citet{lun06}, who considered only $\Sigma_\mathrm{H_2}$ and found an exponent $n = 1.2 \pm 0.2$. Similarly, \citet{sof17} found $n = 1.12 \pm 0.05$ for a set of $(\Sigma_\mathrm{SFR},\Sigma_\mathrm{H_2})$ pairs obtained through binning the Galactic plane in $0.2 \times 0.2$~kpc$^2$ boxes. No clear correlation was found, instead, between $\Sigma_\mathrm{SFR}$ and both $\Sigma_\mathrm{H\textsc{i}}$ and $\Sigma_\mathrm{gas}$. A steeper slope ($3.7 \pm 1.6$), again considering only $\Sigma_\mathrm{H_2}$, was found by \citet{gut11}. However, this value was based only on the analysis of a small sample of nearby star forming regions, which are not necessarily representative of the entire Galaxy.

Since in our case the available data set includes the entire Galactic plane, the approach we follow here is to plot the $(\Sigma_\mathrm{SFR},\Sigma_{\mathrm{H}_2})$ pairs azimuthally averaged in bins of Galactocentric radius (i.e. rings). This was done, for example, in \citet{boi03}, who found that the KS~relation is satisfied in the range $4 \leq R_\mathrm{gal} \leq 15$~kpc with a slope of 2.06, and in \citet{mis06}, who obtained a slope of 2.18 across the entire range of $R_\mathrm{gal}$. 

To do that, for the $x$ axis we use the molecular $\Sigma_{\mathrm{H}_2}$ averaged over Galactocentric bins of 1~kpc by \citet{miv17}, starting from the CO survey assembled by \citet{dam01}, and the atomic $\Sigma_\mathrm{H\textsc{i}}$ from \citet{nak16}. Therefore, for consistency, we recalculated the SFR density in the same bins (which are twice wider than those used for Figure~\ref{radialsfr})\footnote{See \citet{ono10} and \citet{kruj14} for detailed discussions of minimum scales to be preserved while binning, to avoid incomplete sampling and consequent break down of star formation relations.}.

\begin{figure}[t]
\includegraphics[width=0.47\textwidth]{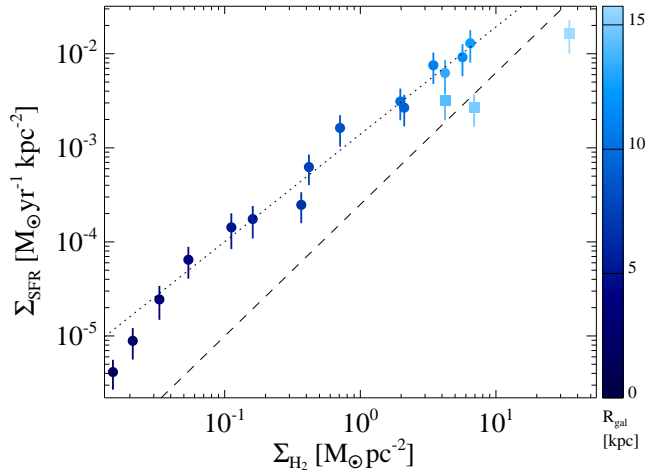}
\caption{Star formation rate (SFR) estimated from the Hi-GAL observations versus the average molecular gas surface density \citep[from][]{miv17} in 1-kpc-wide concentric rings around the Galactic center whose central radius is indicated by the colors.
The bins corresponding to $R_\mathrm{gal} \leq 3 $~kpc, and $R_\mathrm{gal} > 3$~kpc are marked with square and circle symbols, respectively.
The dotted and the dashed lines indicate the linear fit to the SFR estimates at $R_\mathrm{gal} \geq 3$~kpc (with slope $n=\ksn \pm \eksn$), and the KS~relation ($\Sigma_\mathrm{SFR}= 2.5 \times 10^{-4}\, \Sigma_\mathrm{gas}^{1.4}$), respectively.}
\label{kslaw}
\end{figure}

In Figure~\ref{kslaw} the log-log plot of $\Sigma_\mathrm{SFR}$ vs $\Sigma_{\mathrm{H_2}}$ is shown. Differently from the plots built with the atomic and total gas density (see Appendix~\ref{ksappendix}), a linear trend can be recognized here, except for points corresponding to the innermost radii. For this reason, in the estimation of the linear fit, following the indication of \citet{boi03}, we do not consider the points corresponding to three first Galactocentric rings ($R_\mathrm{gal} \leq 3$~kpc). Notice that, as it can be seen in the bottom panel of Figure~\ref{radialsfr}, the fraction of SFR contained within this radius correspond to only the \cumrgalthree\% of the entire Milky Way SFR. From the fit to the remaining points, we obtain $n=\ksn \pm \eksn$.

First of all, from the qualitative point of view, it is important to confirm that the largest part of the Milky Way disk follows a power-law behavior, testifying a direct and clear connection between the availability of molecular gas and the rate of its conversion into stars.

The slope $n$ we found appears in good agreement with those derived by \citet{lun06} and \citet{sof17}, while it is shallower than those found by \citet{boi03} and \citet{mis06}. 
Furthermore, we note the agreement of our result with that obtained by \citet{ono10} for the M33 galaxy ($n=1.18 \pm 0.11$).
The comparison with the classic slope of \citet{ken98}, instead, makes little sense, since it refers to the total $\Sigma_\mathrm{gas}$. 

Actually, it is known that a variety of slopes is generally found for the KS relation \citep[see, e.g.,][and references therein]{ken12}. 
Furthermore, it is not difficult to figure out how varying some assumptions made in calculating the plotted variables can produce a change of the slope (even though preserving the global power-law behavior). For example, surface densities were computed by \citet{miv17} by converting CO to H$_2$ by using the constant factor $X_\mathrm{CO}=10^{20}$~cm$^{-2}~($K~km~s$^{-1})^{-1}$. However, if an increasing trend of $X_\mathrm{CO}$ at increasing $R_\mathrm{gal}$ was adopted \citep[e.g.,][]{nak06,pin13}, the plot in Figure~\ref{kslaw} would get steeper \citep[see also][]{boi03}. 
Moreover, the masses used here as input for the SFR calculation were derived by \citet{eli21} by assuming a constant gas-to-dust ratio of 100. However, a possible increase of this ratio at increasing $R_\mathrm{gal}$ is suggested by \citet{gia17}, and invoked by \citet{eli21} themselves to better explain the Galactocentric behavior of the surface density of Hi-GAL clumps. Taking into account such dependence would produce a flattening of our $\Sigma_\mathrm{SFR}$ vs $\Sigma_{\mathrm{H}_2}$ plot.

\subsection{Face-on view of the Milky Way’s SFR}
Mapping the SFR in the Galactic plane as in Figure~\ref{mapsfr} offers the chance of comparing the local values of this observable with those of other quantities that can be mapped starting from the Hi-GAL clump distribution in the plane.


\begin{figure*}[!t]
\includegraphics[width=\textwidth]{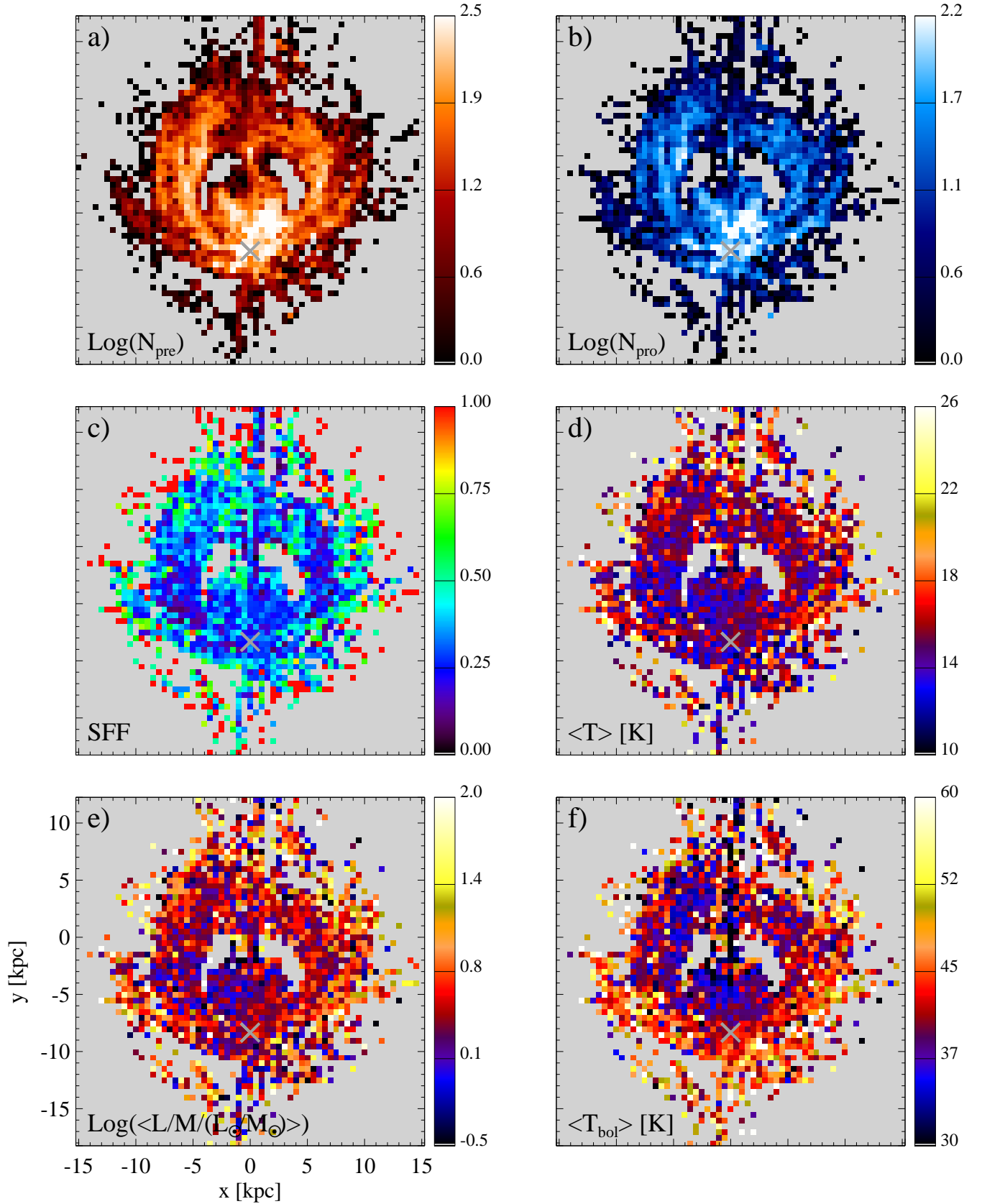}
\caption{Face-on view of the Milky Way representing the following quantities derived (with the same spatial binning used for Figure~\ref{mapsfr}) from the Hi-GAL observations…: $a$) logarithm of number of pre-stellar Hi-GAL clumps per bin, $N_{\rm pre}$; $b$) logarithm of number of protostellar Hi-GAL clumps per bin, $N_{\rm pro}$; $c$) star formation fraction, SFF; $d$) median temperature (from modified black body fit) of protostellar clumps, $\left< T \right>$; $e$) median logarithm of luminosity-to-mass ratio normalized to the solar values, $\left<(L/M)/(\mathrm{L_{\odot}}/\mathrm{M_{\odot}})\right>$; $f$) median bolometric temperature of protostellar clumps,  $\left< T_\mathrm{bol} \right>$. In each panel, the [0,0] position corresponds to the Galactic center position, and the gray X symbol indicates the position of the Sun.
}
\label{npre}
\end{figure*}

In our estimates, the SFR is the sum of the contributions from each protostellar clump, which is a function of its mass.
In this respect, the spatial distribution of SFR in bins of $[x,y]$ coordinates mirrors that of total mass of protostellar clumps in those bins.
This total mass, in turn, does not necessarily depend on the number of protostellar clumps $N_\mathrm{pro}$ found in each bin (shown in Figure~\ref{npre}, $b$)\footnote{Spatial distributions in panels $a$ and $b$ clearly appear asymmetric because at shorter heliocentric distances a larger number of sources (having relatively smaller physical sizes and masses) is detected. Conversely, at higher distances sources tend to appear blended in physically larger structures (Section~\ref{sourcesel}). Since the SFR depends on the available mass (Section~\ref{sfrcalc}), the final appearance of the SFR density map in Figure~\ref{mapsfr} appears much more symmetric than that of source counts. Moreover, the selection bias with distance has to be taken into account; this is done in Appendix~\ref{complappendix}.}, and does not depend at all on the number of pre-stellar clumps $N_\mathrm{pre}$ (shown in Figure~\ref{npre}, $a$). These two quantities were combined together by \citet{rag16} to obtain the star formation fraction (hereafter SFF), defined as the ratio $N_\mathrm{pro}/\left(N_\mathrm{pre}+N_\mathrm{pro}\right)$. The relative populations of these two classes depend on their corresponding lifetimes, which in turn depend on the mass \citep{mot07,urq14c,eli21}. In this respect, the local SFF is related not only to the overall evolutionary state of clumps in a given region, but also, indirectly, to their mass spectrum. Using Hi-GAL data, \citet{rag16} highlighted a decreasing behavior of SFF as a function of the Galactocentric radius over the 3.1~kpc~$< R_\mathrm{gal} < 8.6$~kpc range. This was confirmed by \citet{eli21}, who however found a more scattered behavior just outside this range. This 1-D description of the SFF is further developed here in the 2-D mapping shown in the panel $c$ of Figure~\ref{npre}, in which we can recognize the aforementioned decrease of this quantity at intermediate Galactocentric radii as a blueish ring-like area.

\begin{figure*}[!t]
\includegraphics[width=\textwidth]{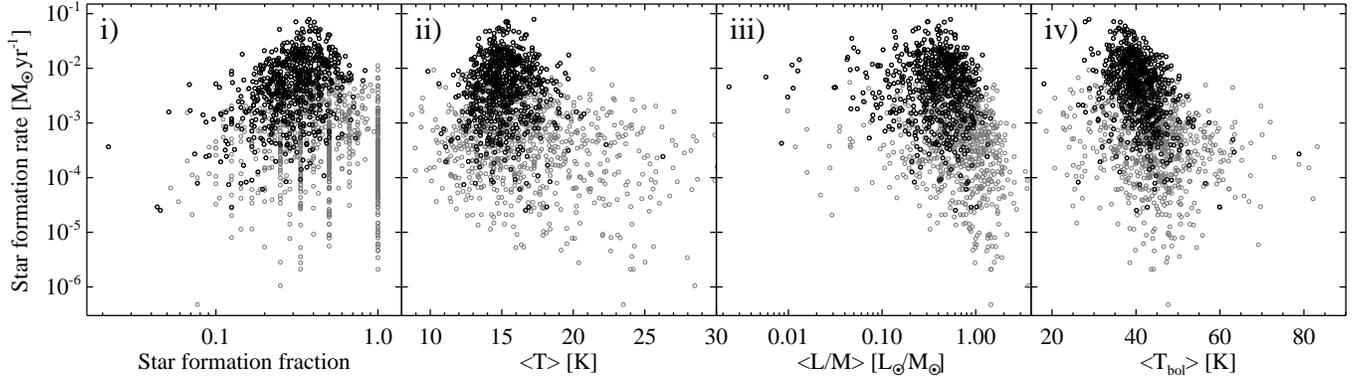}
\caption{Star formation rate (SFR), evaluated in bins of 0.5~kpc in the $[x,y]$ plane, as a function of star formation fraction ($i$), and medians of temperature ($ii$), $L/M$ ratio ($iii$) and bolometric temperature ($iv$) of protostellar clumps, respectively, all calculated in the same spatial bins. In this respect, they represent a pixel-to-pixel plot Figure~\ref{mapsfr} vs panels $c$, $d$, $e$ and $f$ of Figure~\ref{npre}, respectively. Gray symbols correspond to bins with $N_\mathrm{pro}+N_\mathrm{pre}<20$, considered statistically poor. In the leftmost panel, several of them give rise to vertical features corresponding to recurring values like $N_\mathrm{pro}/(N_\mathrm{pro}+N_\mathrm{pre})= \sfrac{1}{1}, \sfrac{1}{2}, \sfrac{1}{3}, \sfrac{1}{4}$, etc.).
}
\label{sfrsfflm}
\end{figure*}

It makes sense to compare here the maps of SFR and SFF, since on the one hand they have no direct link a priori, and on the other hand this allows us to check whether locally the star formation rate is somehow correlated to the average evolutionary stage of the region. A bin-to-bin comparison between SFR and SFF is shown in panel $i$ of Figure~\ref{sfrsfflm}. Apparently, no correlation emerges from the plot, except for the fact that a small number of bins with very low SFF ($\ll 1$) also exhibit relatively low SFR values. 

As a further check of the absence of a trend with the average evolutionary stage of clumps in different parts of the disk, we also investigate the trend of the SFR versus the medians of meaningful evolutionary indicators for protostellar clumps, such as modified black body temperature $T$, bolometric luminosity over mass ratio $L/M$, and bolometric temperature $T_\mathrm{bol}$ \citep{ces15,eli17,eli21}. Median values were evaluated for them in the same spatial bins of Figure~\ref{mapsfr}, and displayed in Figure~\ref{npre}, panels $d$, $e$, and $f$, respectively.

For the median temperature and $L/M$ ratio of protostellar clumps, \citet{eli21} did not find relevant variations in the 1-D profile as a function of $R_\mathrm{gal}$, apart from an increase in the far outer Galaxy, but supported by poor statistics. On the contrary, the trend found for median $T_\mathrm{bol}$ is to slightly raise at increasing $R_\mathrm{gal}$. This is what can be seen also in the 2-D view provided in Figure~\ref{npre}. Therefore, the bin-to-bin comparison of quantities nearly constant with $R_\mathrm{gal}$ such as $T$ and $L_\mathrm{bol}/M$ and an observable with a more complex spatial pattern (SFR) is expected to show no correlation between the two, as confirmed by Figure~\ref{sfrsfflm}, panels $ii$ and $iii$. A mild indication seems to be provided by bins with poor statistics (i.e. containing less than 20 clumps, so an even smaller number of protostellar ones), generally corresponding to large $R_\mathrm{gal}$, for which a higher $L/M$ but a small SFR (due to deficit of clumps) can be found. A similar, but clearer, behavior can be seen also considering the median $T_\mathrm{bol}$ (panel $iv$), essentially due to the increase of this observable with $R_\mathrm{gal}$.

The large degree of scatter between the local SFR and median evolutionary indicators suggests that while the SFR, as it is calculated here, essentially depends on the local availability of mass, it is quite insensitive to the average evolutionary stage of clumps. In fact, it increases with the number of protostellar clumps and with their masses. Therefore, massive star forming regions surely provide a relevant contribution to the SFR of the Milky Way, however this does not seem to depend - once their mass is given - on an earlier or later mean evolutionary stage. This confirms the result found by \citet{kom18} for M~33, namely the fact that the SFR is clearly correlated with the cloud density, in a somehow extended meaning of the KS relation, but not with the evolutionary stage.

\subsection{SFR and spiral arm locations}

In several portions of the spatial distribution of SFR density$\Sigma_\mathrm{SFR}$ in the Galactic plane shown in Figure~\ref{mapsfr} it is possible to recognize stretches of spiral arm features, similarly to what has been observed in external galaxies \citep{reb15,ela19,yu21}.

The SFR arm-like features are not much different from the pattern visible also in the distribution of protostellar clumps (Figure~\ref{npre}, panel~$b$), i.e. the elements from which the SFR is computed. However, as already discussed in \citet{eli17,eli21}, a relevant amount of clumps with ``inter-arm'' distances is also found, due to a large spread in velocities detected along many lines of sight of spectral surveys \citep{meg21}. Additionally, each different analytic spiral arm prescription available in the literature is able to successfully reproduce the observations over wide ranges of longitude, but can fail to fit observed arm-like features along other directions.

\begin{figure}[!h]
\includegraphics[width=8.5cm]{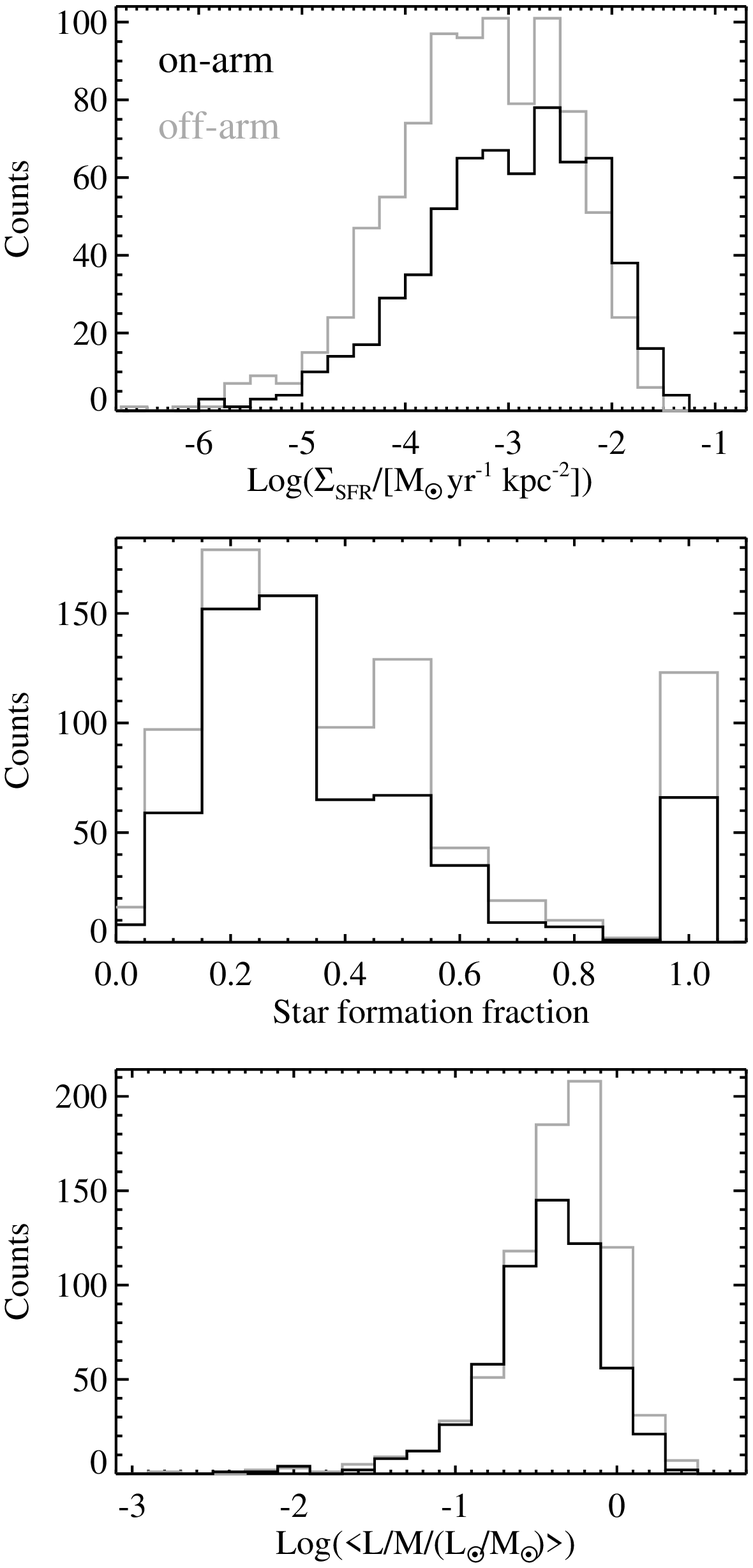}
\caption{\textit{Top}: histograms of logarithm of SFR density for pixels of the map in Figure~\ref{mapsfr}, classified as on-arm (black) and off-arm (gray), based on the fact that their minimum distance from at least one of the spiral arms displayed in Figure~\ref{mapsfr} is shorter or longer than 0.5~kpc, respectively.
\textit{Middle}: the same as in the top panel, but for pixels of the map of SFF in Figure~\ref{npre}, panel $c$.
\textit{Bottom}: the same as in the previous panels, but for pixels of the map of $\left<L_\mathrm{bol}/M\right>$ in Figure~\ref{npre}, panel $e$.
}
\label{spiralhist}
\end{figure} 

In Figure~\ref{mapsfr} we show, as an example, the four-arm prescription by \citet{hou09}. It appears in good agreement with enhancements of our SFR density map along a significant portion of the Perseus arm; a good agreement is also found for the Norma and Carina-Sagittarius arms in the third quadrant, and for the Carina-Sagittarius and Crux-Scutum arms in the first octant of the fourth quadrant, respectively. Nevertheless, there are other regions with high SFR density, but not directly intersected by one of the displayed arms. 
All these aspects are mirrored in the distribution of the $\Sigma_\mathrm{SFR}$ values in Figure~\ref{mapsfr}, if pixels closer than 0.5~kpc from a whatever spiral arm candidate plotted in the figure are considered separately from the remaining ones (Figure~\ref{spiralhist}, top). The distribution of the ``on-arm'' pixels is visibly more left-skewed than the one of the ``off-arm'' pixels.
In short, we observe a sufficiently clear spiral arm structure in the SFR spatial distribution, which elects spiral arms as preferential (although not unique) places for star formation activity in the Galaxy. 

A similar spiral arm-like structure, on the contrary, is not seen in Figure~\ref{npre} in the distributions of both SFF (panel $c$) and median evolutionary indicators as $T$ ($d$), $L_\mathrm{bol}/M$ ($e$), and $T_\textrm{bol}$ ($f$).
Figure~\ref{spiralhist}, middle, contains a comparison between the on-arm and off-arm distributions of SFF. These two distributions appear to be much less distinguishable than in the case of $\Sigma_\mathrm{SFR}$ (top panel). 
A similar behavior can be also seen for $\left<L_\mathrm{bol}/M\right>$ (Figure~\ref{spiralhist}, middle), taken as an example for the three considered evolutionary indicators. 

These considerations reinforce the conclusions of \citet{rag16} and \citet{eli21} about the role of spiral arms, i.e. that no significant differences in the mean evolutionary stage of star formation between spiral arms and inter-arm regions are found. The main difference between these two regions seems rather to consist in the larger amount of material available for star formation in the former than in the latter \citep[see also][]{moo12}.

\section{Summary}\label{summary}
We used the information contained in the catalog of Hi-GAL clump physical properties to obtain a self-consistent estimate the Milky Way SFR.
The main points to be summarized are:
\begin{itemize}
    \item We extrapolated from the algorithm proposed by \citet{ven13} and \citet{ven17} an operative analytic prescription to obtain the contribution to the SFR from the mass of each far-infrared clump: $\textrm{SFR}_\textrm{clump}=\plawfitcoeff \times 10^{\plawfitcoeffexp}~ (M_\textrm{clump}/\textrm{M}_{\odot})^{\plawfitcoexp}~\textrm{M}_{\odot}~\textrm{yr}^{-1}$. 
    \item Considering all clumps classified as protostellar (i.e. star forming) in the catalog and provided with a heliocentric distance, we obtain SFR=$\sfrtot \pm \esfrtot$~M$_{\odot}$~yr$^{-1}$.
    \item 
    
    If we also take into account the contribution from the protostellar clumps without distance assignment, we estimate an additional contribution of \contrsimulave~M$_{\odot}$~yr$^{-1}$ to the global SFR, assuming that their distances follow the same distribution as the sources with known locations.
    This would lead to a final SFR=$\sfrtottot \pm \esfrtottot$~M$_{\odot}$~yr$^{-1}$. Also simulating the extreme case consisting in placing all those sources at $d=13$~kpc, the corrective term would not surpass  \contrsimulthirteenkpc~M$_{\odot}$~yr$^{-1}$.
    \item The profile of SFR surface density as a function of the Galactocentric radius $R_\mathrm{gal}$ is qualitatively similar to other results in the literature for the Milky Way and NGC~6946, and is consistent with most recent models \citep{eva22}. The absolute maximum is found in correspondence of the CMZ, and another local peak is found around $R_\mathrm{gal}=5$~kpc, after which the logarithm of SFR density linearly decreases at increasing $R_\mathrm{gal}$, with slope \slopessfrgal. 
    \item Studying the cumulative of the SFR as a function of the Galactocentric radius, we find that 50\% of the entire Milky Way SFR comes from within the so-called Molecular Ring, and \sfrtotiperc\% from within the Solar circle (inner Galaxy, $R_\mathrm{gal} < 8.34$~kpc). The outer Galaxy is therefore responsible of \sfrtotoperc\% of total SFR and, in particular, the far outer Galaxy ($R_\mathrm{gal} > 13.5$~kpc) of 1\%.
    \item The relation between SFR density and the molecular gas surface density Galactocentric profiles follows, for $R_\mathrm{gal}>3$~kpc (containing the \cumrgalthreecomplement\% of the entire Galactic SFR), a Kennicutt-Schmidt behavior, with slope $n=\ksn \pm \eksn$.
    \item We find no significant trend relating the maps of SFR and those of the observables considered indicative of the mean local evolutionary stage of clumps. We conclude that the local SFR is determined by the amount of available mass rather than by the clump evolutionary stage.
    \item In the distribution of SFR across the Galactic plane, arm-like enhancements emerge over wide ranges of longitude. These enhancements are not seen in maps of the clump evolutionary indicators, suggesting that they are produced by source crowding within the arms.
\end{itemize}

\begin{acknowledgments}
This research has received funding from the INAF Mainstream Grant ``The ultimate exploitation of the Hi-GAL archive and ancillary infrared/mm data'' (1.05.01.86.09), and from the European Research Council synergy grant ECOGAL (855130). SL acknowledges support by the INAF PRIN 2019 grant ONSET. AZ thanks the support of the Institut Universitaire de France.
\end{acknowledgments}

%

\vspace{5mm}
\facilities{Herschel}

\restartappendixnumbering




\appendix

\section{Evaluating the completeness of the total SFR estimate for the Milky Way}\label{complappendix}
Here we provide an estimate of the validity of the total SFR derived for the Milky Way in Section~\ref{sfrcalc} in the light of possible incompleteness of the Hi-GAL catalog for large heliocentric distances. 

\citet{eli17} showed how the mass completeness limit for Hi-GAL, based on the 350-$\mu$m band flux, varies with source temperature and distance. For $T=16$~K, similar to the median temperature for Hi-GAL star forming clumps \citep{eli21}, a distance $d=10$~kpc, and a typical flux completeness limit of 5~Jy for the inner Galaxy \citep{mol16a}, the mass completeness limit amounts to $\sim 90~\textrm{M}_\odot$. One can therefore expect a shortage of detected clumps with mass lower than that, ignored in the SFR calculation. Here we give a quantitative discussion of this possible effect. 


\begin{figure}[h]
\includegraphics[width=8.5cm]{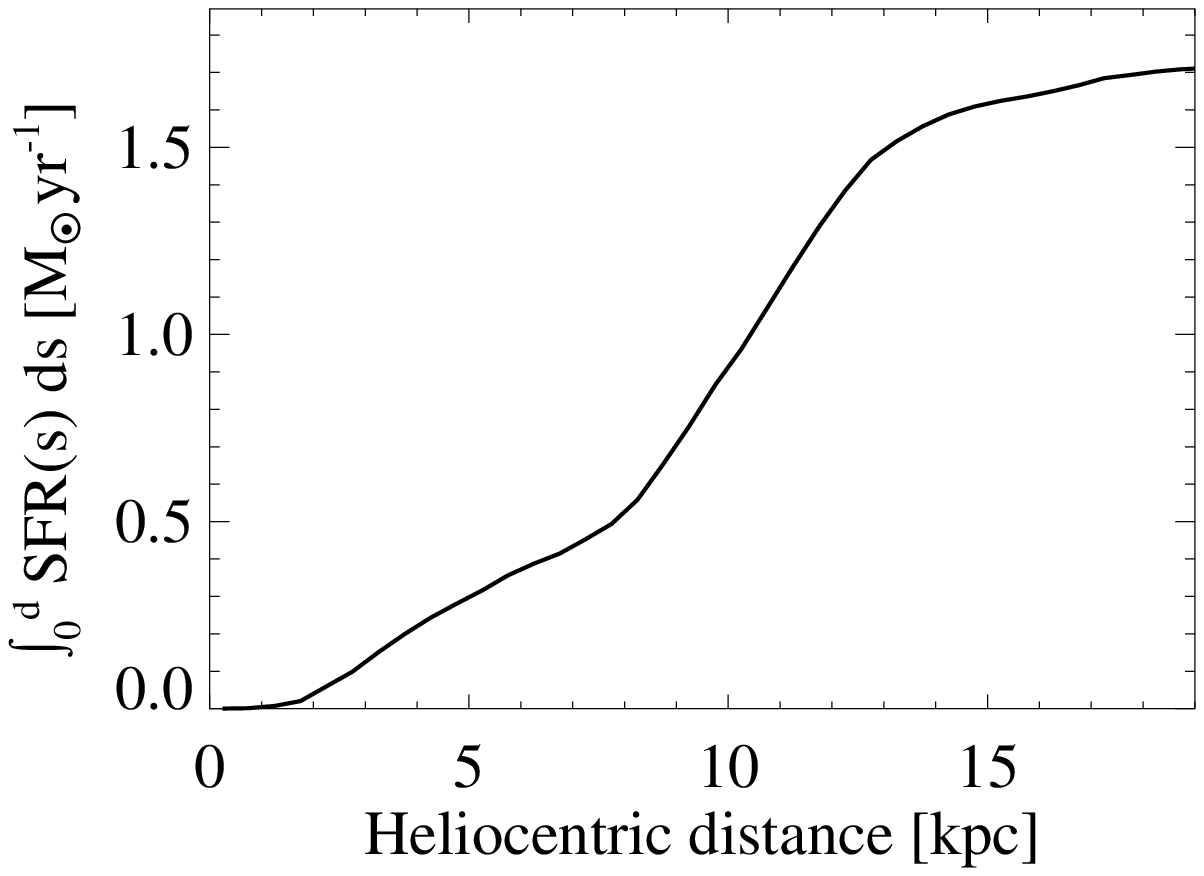}
\centering
\caption{Cumulative SFR profile as a function of the heliocentric distance. Since the distance $d$ appears as upper limit of integration, here the integration variable is called $s$.}
\label{sfrdist}
\end{figure}

First, we compute the cumulative SFR as a function of the distance. In Figure~\ref{sfrdist} a sharp change from a steeply positive to an almost horizontal slope is seen around $d=12$~kpc. The same can be affirmed, but in terms of source number, noting that around the same distance a decreasing trend is seen to start also in Figure~\ref{distdistr}. Such a change of slope can, in principle, be a combination of the true distribution of the SFR density in the plane and the selection effect due to distance we want to quantify. In the Hi-GAL catalog distances $d>12$~kpc are achieved by sources in the first and fourth quadrants\footnote{Notice that, since the relation between Galactocentric radius and heliocentric distance is $R_\mathrm{GC}=\sqrt{R_0^2+d^2+2R_0d\cos\ell}$, sources placed at a same distance span a variety of Galactic locations, with the minimum $R_\mathrm{GC}$ 
achieved for $\ell=0^\circ$.} and the total SFR due to such sources is \sfrmorethantwelve~M$_\odot$~yr$^{-1}$.

Second, we evaluated the SFR from the symmetrical region on the opposite side of the Milky Way, i.e. we considered clumps (essentially located in the second and third quadrants) lying outside the circle centered on the symmetric point of the Sun with respect to the Galactic center ($[x,y]=[0,+8.34]$~kpc) and with a radius of 12~kpc. The SFR from this area amounts to \sfrmorethantwelveopposite~M$_\odot$~yr$^{-1}$, so that, in the rough assumption of a circular symmetry for the  ``true'' SFR distribution in the $[x,y]$ plane, the  ``missing SFR'' due to incompleteness is comparable to the difference in SFRs measured from the two zones, namely \sfrmorethantwelvediff~M$_\odot$~yr$^{-1}$. The extent of this correction shows that the effect of the catalog incompleteness is not much strong, although not negligible: it should be added as a further term to the total estimate of \sfrtot~M$_\odot$~yr$^{-1}$ based on clumps provided with a distance estimate (Section~\ref{sfrcalc}), then representing a \sfrmorethantwelveperc\% correction to it.

To evaluate an analogous additional term for the SFR of \sfrtottot~M$_\odot$~yr$^{-1}$, estimated by involving also clumps with no distances (Section~\ref{nodist}) is more complicated, because the distribution itself of known distances is involved in the calculation. However, assuming a \sfrmorethantwelveperc\% correction also in this case, the total Milky Way SFR would amount to \sfrtottotcorrectedmorethantwelve~M$_\odot$~yr$^{-1}$.

\section{Testing the Kennicutt-Schmidt relation with atomic gas}\label{ksappendix}
\begin{figure}[t]
\includegraphics[width=17cm]{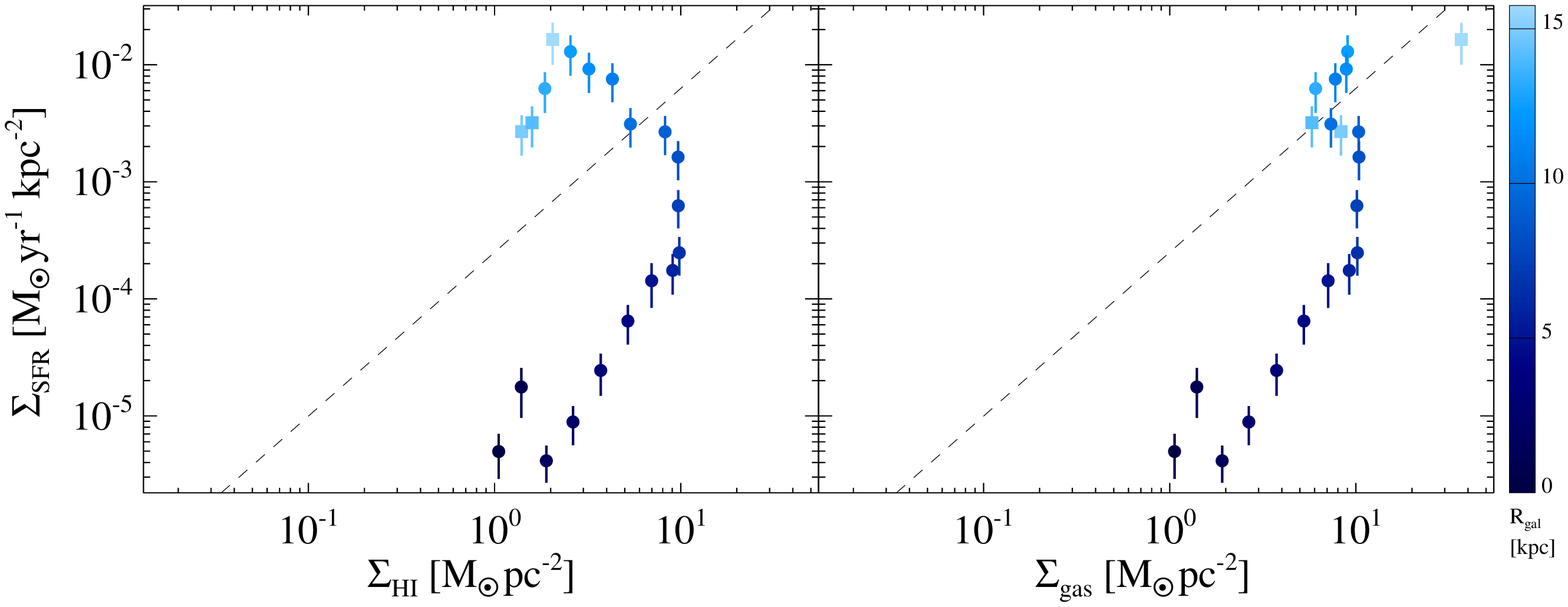}
\centering
\caption{\textit{Left}: the same as in Figure~\ref{kslaw}, but with atomic gas surface density on the $x$ axis \citep[from][]{nak16}, instead of the molecular one.
The KS relation is also plotted as a dashed line. \textit{Right}: the same as in Figure~\ref{kslaw} and in the left panel of this figure, but with total gas surface density on the $x$ axis. In both panels the $x$-axis range has been kept identical to that of Figure~\ref{kslaw}, to facilitate comparison.}
\label{ksappendixfig}
\end{figure}

To complete the discussion about the SFR vs gas density relation contained in Section~\ref{kssection}, here we show the plots of $\Sigma_\mathrm{SFR}$ vs $\Sigma_\mathrm{H\textsc{i}}$ (Figure~\ref{ksappendixfig}, left) and $\Sigma_\mathrm{gas}=\Sigma_\mathrm{H\textsc{i}}+\Sigma_{\mathrm{H}_2}$ (right), respectively, obtained using the Galactocentric profile of H\textsc{i} surface density of \citet[their Figure~4]{nak16}.

Comparing such profile with that of $\Sigma_\mathrm{SFR}$, peaks for these two quantities are seen at very different values of $R_\mathrm{GC}$ (at $\sim 10$ and $\sim 5$~kpc, respectively), preventing a power-law relation between the two. Furthermore, for $R_\mathrm{GC} \lesssim 20$~kpc, $\Sigma_\mathrm{H\textsc{i}}$ varies only within one order of magnitude (namely between $\sim 1$ and 10~M$_\odot$~pc$^{-2}$), so that the plot assumes an almost vertical appearance, as already highlighted by \citet{sof17} and \citet{bac19}. In this last work, in particular, it was shown how a power law behavior is recovered if the atomic+molecular gas volume density instead of the surface density is considered, which implies to take into account the increasing Galactic disk scale height at increasing $R_\mathrm{GC}$.

Similar considerations can be formulated if one considers also the total gas surface density, which, as it can be seen in \citet{miv17}, is dominated by the atomic component at $R_\mathrm{GC} \gtrsim 7$~kpc.

\bibliographystyle{apj}
\bibliography{SFR}


\end{document}